\documentclass[12pt,a4paper]{article}
\pdfoutput=1
\usepackage{jheppub}
 \usepackage{amsmath}
 \usepackage{amsfonts}
 \usepackage{mathtools}
 \usepackage{amssymb}
 \usepackage{caption}
 \usepackage{subcaption}
 \usepackage{slashed}
 
  \title{Bootstrap and collider physics of parity violating conformal field theories in $d=3$}
  \author{Subham Dutta Chowdhury ${}^{a,c}$, Justin R. David ${}^a$, Shiroman Prakash ${}^b$}
\affiliation{ ${}^a$ Centre for High Energy Physics, Indian Institute of Science,\\
C. V. Raman Avenue, Bangalore 560012, India.\\
${}^b$Department of Physics and Computer Science, \\
Dayalbagh Educational Institute, Dayalbagh, \\
Agra 282005, India.\\
${}^c$Tata Institute of Fundamental Research,\\
Homi Bhabha Road, 
Mumbai 400005, India.
}
\emailAdd{subham@theory.tifr.res.in, justin@iisc.ac.in, shiroman@gmail.com}
\abstract{ We study  the crossing equations in $d=3$  for the four point function
of two $U(1)$ currents and two scalars including  the presence of a parity violating term for the 
$s$-channel stress tensor exchange. We show the existence of a new tower of double trace operators in the $t$-channel whose presence is necessary for the crossing equation to be satisfied and determine the corresponding large spin  parity violating OPE coefficients.  Contrary to the parity even situation, we find that  the parity odd  
 $s$-channel  light cone stress tensor  block 
  do not have logarithmic singularities. This implies that the parity odd term 
does not contribute to 
 anomalous dimensions in the crossed channel at this order light cone expansion. 
We then study 
the constraints imposed by reflection positivity and crossing symmetry on such a four point function. We reproduce the previously known parity odd collider bounds  through this analysis. 
The contribution of the parity violating term in the collider bound results from 
 a square root branch cut present in the light cone block as opposed to a logarithmic cut in the parity even case, 
  together with  
the application of the Cauchy-Schwarz inequality.    
}
 
\begin{document}

\maketitle
\section{Introduction}
\begin{figure}[h]
\center
\includegraphics[scale=0.7]{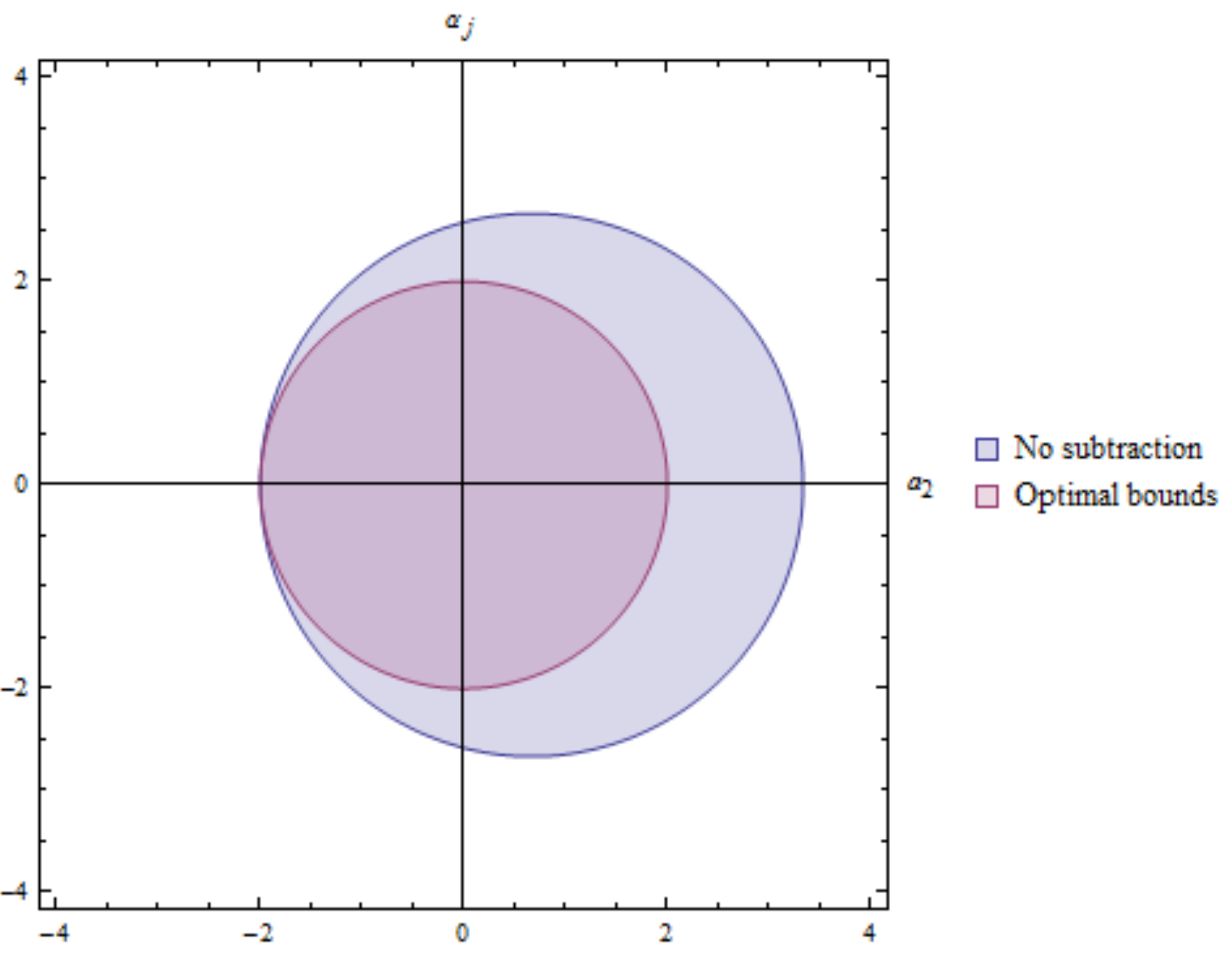}
\caption{Comparative study of bounds}
\label{first}
\end{figure}

\paragraph{ } Conformal field theories (CFT) have been very well studied, particularly in $d=3$, where they have been central to the understanding of several important phenomena in both condensed matter physics and holography \cite{Maldacena:1997re, Klebanov:2002ja,Leigh:2003gk,Sezgin:2003pt,Aharony:2008ug,Aharony:2008gk}. The three dimensional CFTs are special in the sense that presence of a Chern-Simons term means that we can consider theories which violate parity \cite{Giombi:2011kc,Aharony:2011jz}. Conformal invariance then completely fixes the three point functions of operators, two $U(1)$ currents and a stress tensor, for example, in such theories upto two parity even and one parity odd independent parameters. Consider the three point function of two conserved $U(1)$ currents and a conserved stress tensor,
\begin{eqnarray}
\langle jjT \rangle &=& n^j_s\langle jjT \rangle_{\textrm{free boson}} 
+n^j_{f}\langle jjT \rangle_{\textrm{free fermion}} + p_j \langle jjT \rangle_{\textrm{parity odd}},\nonumber  
\end{eqnarray}
where  $\langle .. \rangle_{\textrm{free boson}}$, 
$\langle .. \rangle_{\textrm{free fermion}}$ denote the correlator  a real free  boson and a real free fermion respectively \cite{Osborn:1993cr}, while $\langle .. \rangle_{\textrm{parity odd}}$ refers to the parity odd structure \cite{Giombi:2011rz}. The  numerical coefficients  $n_s^{j}, n_f^{j}$ correspond to the parity invariant sector, while $p_{j}$ is the parity violating coefficient. The parity violating structure is unique to $d=3$ and it  appears only for interacting theories. \paragraph{ } Constraints on the parameter space of three point functions of CFTs were first studied for $d=4$ in \cite{Hofman:2008ar}. The authors consider localized perturbations of CFT in Minkowski space which spread out in time.  The energy measured in a direction, denoted by $\hat{n}$, is defined by,
\begin{eqnarray}
E_{\hat n} &=& \lim_{r \rightarrow \infty} r \int_{-\infty}^{\infty} dt~ n^iT^{t}_i(t, r \hat n),
\end{eqnarray}
where $r$ is the radius of the circle on which the detector is placed and $\hat n$ is a unit vector which determines the point on the circle where the detector is placed. The expectation value of the energy flux measured in such a way should be positive for any state. This led to constraints on parameters of three point functions $\langle TTT \rangle$ and $\langle JJT \rangle$ \cite{Hofman:2008ar}. This analysis was generalized to higher dimensions in \cite{Buchel:2009sk, Myers:2010jv}. The central assumption to the collider physics analysis is the average null energy is positive over any state. 
\paragraph{}Particularly, for $d=3$, the collider bounds constrain the parameters to lie within a circle on the $(\alpha_j - a_2)$ plane \citep{Chowdhury:2017vel}. 
\begin{eqnarray}\label{optibo}
&&\alpha_j^2 + a_2^2 \leq  4,\nonumber\\
&& a_2 =-\frac{2 (n^j_f-n^j_s)}{(n^j_f+n^j_s)} ,\qquad
\alpha_j = \frac{4 \pi ^4 p_j}{(n^j_f+n^j_s)}.
\end{eqnarray}
Chern-Simons theories coupled to fundamental matter were shown to saturate this bound, that is they lie at the boundary of the disc
 \cite{Chowdhury:2017vel, Cordova:2017zej}. Note that there are always certain assumptions (namely the fact that positivity of average null energy being satisfied in an interacting CFT) beyond the basic assumptions of CFTs. 
\paragraph{ }
             On the other hand, causality considerations of the Lorentz invariant Minkowski CFT can lead to non-trivial constraints on the OPE data of the CFT \citep{Hartman:2015lfa, Hartman:2016dxc, Hartman:2016lgu,Afkhami-Jeddi:2017rmx, Afkhami-Jeddi:2016ntf,Afkhami-Jeddi:2018apj,Meltzer:2018tnm,Afkhami-Jeddi:2018own,Meltzer:2017rtf}. In \citep{Hartman:2015lfa}, the authors consider a four point function of scalar operators. The crux of the argument relied on the  fact that well-behaved Euclidean theories are in one to one correspondence with causal Minkowski theories. Starting from an Euclidean correlator, one gets all possible Lorentzian correlators with desired time-orderings by analytic continuations. Singularities of the Euclidean theories continue to be the singularities of the Minkowski theories but now there are branch cuts appearing in the Minkowski theory exactly at the light cone. A diagnostic of a causal Minkowski theory is the fact that commutators of operators must be vanishing at space-like separations. 
\begin{eqnarray}
\langle \psi| [\mathcal{O}(x), \mathcal{O}(0)] |\psi \rangle = 0, \qquad x^2 \geq 0  
\end{eqnarray}    
\indent
It is easy to see that this commutator becomes non-zero when one operator crosses the light cone of the other and not before that. This imposes restrictions on the behaviour of the analytic continuations of the Euclidean four point function  $\langle \psi (0) \mathcal{O} (z, \bar{z}) \mathcal{O} (1,1) \psi(\infty)\rangle$. In short, reflection positivity and crossing symmetry ensures that certain Minkowski correlators, often referred to correlators on the second sheet  are analytic within a specific region of the complex $(z, \bar{z})$ plane and cannot grow faster than the Euclidean correlator (correlator on the first sheet). The resulting sum rules constrains the product of OPE coefficients appearing in the light cone expansion of the aforementioned correlator. This was generalized to stress tensor exchange corresponding to spinning correlators in \cite{Hofman:2016awc}. The light cone limit of spinning correlator for stress tensor exchange is obtained using the differential operator formalism outlined in \cite{Costa:2011dw}. Analyticity and reflection positivity properties of the Euclidean correlator were used to obtain constraints on the parity even parameters of the three point functions $\langle JJT \rangle$ and $\langle TTT \rangle$ in $d\geq 3$. The analysis of the parity even parameters of the spinning correlators involved a non-trivial bootstrap decomposition of the correlator into sum over composite operators in the dual channel to reproduce optimal bounds. The relationship between causality analysis of the Lorentzian theory using bootstrap arguments and average null energy condition, as used by Maldacena et. al. was understood in \cite{Hartman:2016lgu}. Any Lorentz invariant unitary quantum field theory obeys the average null energy condition. The collider bounds are a direct consequence of this. This also produces the optimal bounds for the spinning correlators without having to resort to any subtractions  as done in \cite{Hofman:2016awc}.  The method 
however bypasses  solving the crossing equations or the anomalous dimensions of the high-spin composite operators in the dual channel.
\paragraph{} 
             In this paper we study the modifications to the crossing equation due to the presence of the parity violating terms in the three point functions of $\langle JJT \rangle$ via light cone bootstrap \cite{Li:2015itl} and extend the causality arguments of \citep{Hartman:2015lfa, Hofman:2016awc} to parity violating theories in $d=3$.  We consider the following four point function 
\begin{eqnarray}
\langle J(P_1,Z_1)J(P_2,Z_2) \phi(P_3,Z_3)\phi(P_4,Z_4) \rangle .
\end{eqnarray}
Here  the $P, Z$ refer to the positions and polarizations in the embedding formalism. 
We find  the crossing equation  satisfied by the above correlator in the light cone limit requires additional 
terms due 
to the presence of the parity violating exchange in the $s$-channel conformal block. Crossing symmetry of the four point function of two $U(1)$ currents and two scalars implies
\begin{eqnarray}
&&C_J \frac{H_{12}}{P_{12}^3 P_{34}^{\Delta_\phi}} + \frac{\lambda_{\phi\phi T}}{ \sqrt{C_T}} (D_{\text{even}} - D_{\text{odd}}) {\cal W}(2, 2, \Delta_\phi, \Delta_\phi) =\nonumber\\
&& \sum_{\tau, l} P_{[j,\phi]_{\tau,l}} D^t_{[j,\phi]_{\tau,l}} {\cal W}_{[j,\phi]_{\tau,l}}^t(2, 2, \Delta_\phi, \Delta_\phi)+ P_{\tilde{[j,\phi]}_{\tau,l}} D^t_{\tilde{[j,\phi]}_{\tau,l}} {\cal W}_{\tilde{[j,\phi]}_{\tau,l}}^t(2, 2, \Delta_\phi, \Delta_\phi)\nonumber\\
&&+ P^{11}_{m,\tau,l} D^{11}_m {\cal W}_{\tilde{\mathcal{O}}}^t(2, 2, \Delta_\phi, \Delta_\phi) +P^{12}_{m,\tau,l} D^{12}_m {\cal W}_{\tilde{\mathcal{O}}}^t(2, 2, \Delta_\phi, \Delta_\phi) \nonumber\\
&&+ P^{21}_{m,\tau,l} D^{21}_m {\cal W}_{\tilde{\mathcal{O}}}^t(2, 2, \Delta_\phi, \Delta_\phi) +P^{22}_{m,\tau,l} D^{22}_m {\cal W}_{\tilde{\mathcal{O}}}^t(2, 2, \Delta_\phi, \Delta_\phi), \nonumber\\ \label{bacro}
\end{eqnarray}
where \begin{equation} \label{defcj}
C_J = \frac{n_f^j + n_s^j}{16\pi^2},
\end{equation}
and ${\cal W}$ is the scalar block, the operator in the subscript refer to the operator exchanged in the 
channel. On 
on the RHS the blocks are given  by eqns \eqref{spinningblocksschannel} and \eqref{scblock} and action of the differential operators $D_{\text{even}}$ and $D_{\text{odd}}$ are given in eqns \eqref{deven} and \eqref{dodd} respectively. Their action on the scalar block is detailed in the Appendix \ref{lhsdiffops}. The differential operators $D^t_{[j,\phi]_{\tau,l}}$ and $ D^t_{\tilde{[j,\phi]}_{\tau,l}}$ were already known  in \cite{Li:2015itl} , see equation  \eqref{parityevenoddbuildingblocksrhs}.  On the left hand side ${\cal W}$ refers to the scalar block with the 
stress  tensor exchange. 
 Note that due to the presence of the parity violating term on the LHS
of (\ref{bacro}) for the 
stress tensor exchange, 
we need to modify the RHS of the crossing equation with the differential operators $D_m^{ij}$. The explicit expressions for the operators are given in equation \eqref{tcblocksmixed}. The parity odd contribution is fundamentally different from the parity even contribution. While the parity even contribution to the bootstrap equation on the LHS, is responsible for the anomalous dimensions of the composite operator and corrections to the OPE coefficients $P_{[j,\phi]_{\tau,l}}$ and$P_{\tilde{[j,\phi]}_{\tau,l}}$ on the RHS \cite{Li:2015itl}. 
 The parity odd term on the LHS  results in 
  the new 
   OPE coefficients $P^{ij}_{m,\tau,l}$.
    In a sense the contribution of the parity odd terms on the LHS is on the same footing as the identity contribution\footnote{The fact that parity odd terms might not contribute to the anomalous dimensions was also observed by the authors of \citep{Sleight:2018ryu,Sleight:2018epi}.}. We find 
\begin{eqnarray}
P^{11}_{m,\tau,l} &=& -P^{21}_{m,\tau,l}=-\frac{\sqrt{\pi } 2^{-\Delta_\phi+1} ip_j \lambda_{\phi\phi T}}{\Gamma \left(\Delta_\phi-\frac{1}{2}\right)}\frac{l^{(\Delta_\phi-3)}}{\sqrt{C_T}2^{l}}, \nonumber\\
P^{12}_{m,\tau,l} &=& -P^{22}_{m,\tau,l}=\frac{\sqrt{\pi } 2^{-\Delta_\phi+1} ip_j\lambda_{\phi\phi T}}{\Gamma \left(\Delta_\phi-\frac{1}{2}\right)}\frac{l^{(\Delta_\phi-4)}}{\sqrt{C_T} 2^{l}}.
\end{eqnarray}
We note that the absence of the anomalous dimensions is due to the fact that while, under the action of the operators $D_{\text{even}}$, the resulting block has a logarithm, and hence is responsible for anomalous dimensions in the $t$-channel double trace operators.  The blocks under  the action 
$D_{\text{odd}}$ are rational functions. This is due to the asymmetric shifts of dimensions in 
the differential operators $D_{\text{odd}}$ compared to the symmetric shift of the dimensions in 
$D_{\text{odd}}$\footnote{See equations eqn \eqref{dodd} and \eqref{deven}.}. 
To  be more definite,   let us examine  the parity even light cone block, i.e in the limit $u \ll v$  for $d=3$ 
and 
 the parity odd block, 
  before explicitly acting on by the differential operators but after the shifts in the dimensions
of the operators have been performed. 
   
\begin{eqnarray}\label{introexample}
G_{d=3}^{\text{even}}(u,v) &=& \frac{1}{4} \sqrt{u} ( v-1)^2 \, {}_2F_1\left( \frac{5}{2}, \frac{5}{2},  5; 1-v \right), 
\nonumber \\
%
G_{d=3}^{\text{odd}}(u,v)&=&\frac{4 \sqrt{u} (1-v)^2}{\left(\sqrt{v}+1\right)^4 \sqrt{v}}.
\end{eqnarray}     
In the limit,  $u \rightarrow 0, v\rightarrow 0$, due to the absence of the log terms in the parity odd conformal blocks, this does not contribute to the anomalous dimensions of composite operators in the dual channel. We observe the absence of anomalous dimensions in the leading order solution of the conformal blocks in the light cone limit. Since our analysis of the parity odd blocks is at the leading order in the light cone limit, 
this observation of the absence of the contribution of the parity odd terms to the anomalous 
is true at the leading order in the light cone expansion. 
It  will be  interesting to study whether sub leading terms in the light cone expansion 
of the parity odd term contribute  to the anomalous dimensions in the 
corresponding double trace operators \footnote{We thank Tom Hartman for raising this point.}.
\paragraph{ }             We then consider the constraints imposed by reflection positivity and crossing symmetry on the four point functions of two $U(1)$ currents and two scalars. 
We show that the parity odd term contributes to the conformal collider bound through 
 the Cauchy-Schwarz inequality. The  two bounds we obtain  can be summarized in the figure \ref{first} below
             
 The bounds obtained using the naive 
  the bootstrap analysis is not optimal, it results in the larger circle in figure \ref{first}. 
  The optimal bound  given in (\ref{optibo}) which agrees with that from the average null energy condition
  results by appropriately isolating the contribution of the two  classes of composite operators 
  labelled by different twists 
 as done earlier for parity even theories by 
   \citep{Hofman:2016awc}. 
   Furthermore we 
   note that the nature of the singularity of the light cone limit of the Euclidean parity odd correlator on the second sheet is different from that of the parity even one. To make it more precise, let us consider the light cone blocks corresponding to the party even and odd OPE coefficients before applying the differential operators given in \eqref{introexample}. Conformal invariance allows us to choose the following kinematics for our four point function
\begin{equation}
\langle \epsilon.J(0) \phi(z, \bar{z})\phi(1,1)\epsilon.J(\infty) \rangle .
\end{equation}
Under the analytic continuation $z \rightarrow ze^{-2i\pi}$, both the blocks develop singularities on the second sheet. While, for the parity even case, the origin of this singularity is due to the  presence of  the logarithm  term, for the parity odd case  such a term is absent but nevertheless the square roots in the denominator  of \eqref{introexample}
are responsible for the enhanced singularity on the second sheet.
\paragraph{} The organization is as follows. In section \ref{parityoddthreeptfn} and \ref{jjphiphifromphi}, we set up the differential operator which correctly reproduces the contribution of the stress tensor block in the light cone limit.  The modifications to the crossing equation is discussed in section \ref{opeandanomalousdimensions}. In section \ref{positivityconstraints}, we compute the causality constraints on the four point function of two $U(1)$ currents and two scalars.
    
 \section{Spinning Correlator: $JJT$ from $\phi \phi T$}\label{parityoddthreeptfn}
           \paragraph{ } In this section we  derive the parity odd contribution to the 
            three point correlation function involving two $U(1)$ currents, $j$, of scaling dimension 2 ($\Delta=2$) and a stress tensor $T$ ($\Delta = 3$ ) \citep{Giombi:2011rz} from the three point function of the stress tensor with two scalar operators $\phi$ with the same scaling dimension of the currents in $d=3$ \cite{Osborn:1993cr}. Following \citep{Costa:2011dw}, the differential operator required to obtain the parity odd $JJT$ correlator from $\phi \phi T$ is given by,
\begin{eqnarray}\label{differentialop}
\langle j(P_1;Z_1) j(P_2;Z_2) T(P_3;Z_3) &=& \left(\alpha {\cal D}^{(3)}_{\textrm{left}} + \beta {\cal D}^{(4)}_{\textrm{left}}\right) \langle \phi(P_1;Z_1) \phi(P_2;Z_2) T(P_3;Z_3) \rangle, \nonumber\\
\end{eqnarray} 
where we have expressed the correlators in the embedding space formalism.
\begin{eqnarray}
{\cal D}^{(3)}_{\textrm{left}}\langle \phi(P_1;Z_1) \phi(P_2;Z_2) T(P_3;Z_3) \rangle &=& \left(8 \left\lbrace   \tilde{1}\right\rbrace + 4\left\lbrace   \tilde{2}\right\rbrace\right),  \nonumber\\  
{\cal D}^{(4)}_{\textrm{left}}\langle \phi(P_1;Z_1) \phi(P_2;Z_2) T(P_3;Z_3) \rangle &=& \left(8 \left\lbrace   \tilde{3}\right\rbrace + 4\left\lbrace   \tilde{4}\right\rbrace\right),  \nonumber\\
\end{eqnarray}
where $\left\lbrace \tilde{i} \right\rbrace$ is the differential basis and is given by
\citep{Costa:2011dw}
\begin{eqnarray}
\left\lbrace \tilde{1} \right\rbrace &=& \frac{-6 \epsilon_{13} H_{23} +2 \epsilon_{23}H_{13} -4 \epsilon_{23} V_{1,23} V_{3,12}}{64(P_1.P_2) (P_1.P_3)^3 (P_2.P_3)^3},\nonumber\\
\left\lbrace \tilde{2} \right\rbrace &=& \frac{10 \epsilon_{13} H_{23} -6 \epsilon_{23} H_{13} +4 V_{3,12} (5 \epsilon_{13} V_{2,31}-2 \epsilon_{23} V_{1,23})}{64(P_1.P_2) (P_1.P_3)^3 (P_2.P_3)^3},\nonumber\\
\left\lbrace \tilde{3} \right\rbrace &=& \frac{-2 \epsilon_{13} H_{23} +6 \epsilon_{23} H_{13}+4 \epsilon_{13} V_{2,31} V_{3,12}}{64(P_1.P_2) (P_1.P_3)^3 (P_2.P_3)^3},\nonumber\\
\left\lbrace \tilde{4} \right\rbrace &=& \frac{-10 \epsilon_{23} (H_{1,3}+2 V_{1,23} V_{3,12})+6 \epsilon_{13} H_{23}+8 \epsilon_{13} V_{2,31} V_{3,12}}{64(P_1.P_2) (P_1.P_3)^3 (P_2.P_3)^3},
\end{eqnarray}
where
\begin{eqnarray}\label{hvdefn}
H_{ij} &=& -2\left[ (Z_i.Z_j)(P_i.P_j) - (Z_i.P_j)(P_i.Z_j) \right], \nonumber\\
V_{i,jk} &=& \frac{(Z_i.P_j)(P_i.P_k)-(Z_i.P_k)(P_i.P_j)}{(P_j.P_k)},\nonumber\\
\epsilon_{ij} &=& (P_i.P_j) \epsilon (Z_i,Z_j,P_1,P_2,P_3).  
\end{eqnarray}
Note that $P_i$s and $Z_i$s are defined in a five dimensional embedding space. To project into the real 3 dimensional space, one uses the following projection formulae
\begin{eqnarray}\label{projection}
P^\mu_i &=& (1, x_i^2, x_i^\mu),\nonumber\\
Z^\mu_i &=& (0, 2x_i.z_i, z^\mu_i) ,\nonumber\\
P_i.P_j &\rightarrow & -\frac{1}{2}x_{ij}^2, \nonumber\\
P_i.Z_j &\rightarrow & z_j.x_{ij}, \nonumber\\
Z_i.Z_j &\rightarrow & z_i.z_j, \nonumber\\
\epsilon (Z_i,Z_j,P_1,P_2,P_3)&\rightarrow & \begin{vmatrix}
0 & 0 & 1 & 1 & 1  \\
2x_i.z_i & 2x_j.z_j & x_1^2 & x_2^2 & x_3^2 \\
z_i^\mu & z_j^\nu & x_1^\rho & x_2^\sigma & x_3^\gamma\\
\end{vmatrix}.
\end{eqnarray}
In order to compare it against the known structure of parity odd three point functions in literature \citep{Giombi:2011rz, Giombi:2011kc} we re-express the building blocks of the three point functions in the embedding space i.e $H_{ij}$, $V_{i,jk}$ and $\epsilon_{ij}$, in terms of the tensor structures used in \citep{Giombi:2011kc}
\begin{eqnarray}\label{transformation}
H_{12} & \equiv & -2x_{12}^4 \tilde{P}_1^2,\nonumber\\
H_{13} & \equiv & -2x_{13}^4 \tilde{P}_2^2,\nonumber\\
H_{23} & \equiv & -2x_{23}^4 \tilde{P}_3^2,\nonumber\\
V_{1,23} & \equiv & -\frac{x_{12}^2x_{31}^2}{x_{23}^2} \tilde{Q}_3,\nonumber\\
V_{2,31} & \equiv & -\frac{x_{12}^2x_{23}^2}{x_{31}^2} \tilde{Q}_2,\nonumber\\
V_{3,12} & \equiv & -\frac{x_{23}^2x_{31}^2}{x_{12}^2} \tilde{Q}_1 ,\nonumber\\
\epsilon_{13} & \equiv & 4 \frac{\tilde{S}_2 (|x_{23}||x_{12}||x_{31}|^5)}{i} ,\nonumber\\
\epsilon_{23} & \equiv & -4 \frac{\tilde{S}_3 (|x_{23}|^5|x_{12}||x_{31}|)}{i},\nonumber\\
\epsilon_{12} & \equiv & -4 \frac{\tilde{S}_1 (|x_{23}||x_{12}|^5|x_{31}|)}{i}, 
\end{eqnarray}
where $P_i$, $Q_i$ and $S_i$
are defined in \citep{Giombi:2011rz}. They are related to $\tilde{P}_i$, $\tilde{Q}_i$ and $\tilde{S}_i$ in the following manner,
\begin{eqnarray}
\tilde{P}_i^2 &=& P_i^2|_{1\leftrightarrow 3}.  
\end{eqnarray}
The three point function defined in \citep{Chowdhury:2017vel} is given by,
\begin{eqnarray}\label{shiroman}
\langle jjT \rangle &=&  p_j\frac{\tilde{Q}_1^2\tilde{S}_1+2\tilde{P}_2^2\tilde{S}_3+2\tilde{P}_3^2\tilde{S}_2}{|x_{12}|x_{23}|x_{31}|}.\nonumber\\
\end{eqnarray}
Using the identities in equations
 \eqref{transformation} and \eqref{projection} in equation \eqref{differentialop}, we compare it with \eqref{shiroman}. We use the following identity in getting rid of $\tilde{S}_1$ in eqn \eqref{shiroman}  
\begin{eqnarray}
\tilde{Q}_1^2\tilde{S}_1 &=& (-2\tilde{P}_2^2+\tilde{Q}_1\tilde{Q}_3) \tilde{S}_3 + (-2\tilde{P}_3^2+\tilde{Q}_1\tilde{Q}_2) \tilde{S}_2.
\end{eqnarray} 
We find that,
\begin{eqnarray}
\langle j(P_1;Z_1) j(P_2;Z_2) T(P_3;Z_3) &=&\frac{-i p_j }{1152} \left({\cal D}^{(3)}_{\textrm{left}} + {\cal D}^{(4)}_{\textrm{left}}\right) \langle \phi(P_1;Z_1) \phi(P_2;Z_2) T(P_3;Z_3) \rangle .\nonumber\\
\end{eqnarray}
\section{Spinning conformal blocks and crossing symmetry}\label{jjphiphifromphi}
\paragraph{ } The spinning conformal blocks for two $U(1)$ currents and two scalars can be formally written down as follows,
\begin{eqnarray}
H(z,\bar{z}) &=& \langle j(P_1,Z_1)j(P_2,Z_2) \phi(P_3) \phi(P_4) \rangle,\nonumber\\
&=&\left( \frac{P_2.P_4}{P_1.P_4} \right)^{\frac{1}{2}\sigma_{12}}  \left( \frac{P_1.P_4}{P_1.P_3} \right)^{\frac{1}{2}\sigma_{34}} \sum_{{\mathcal{O}}}\lambda_{jj{\mathcal{O}}}\lambda_{\phi\phi{\mathcal{O}}}\sum_k\frac{Q_{k}(Z_i,P_i)G^{\Delta_1,\Delta_2,\Delta_3,\Delta_4}_{{\cal O},k}(u,w)}{(-2P_1.P_2)^{\frac{1}{2}(\sigma_1+\sigma_2)}(-2P_3.P_4)^{\frac{1}{2}(\sigma_3+\sigma_4)}}, \nonumber\\
u &=& \frac{P_1.P_2 P_3.P_4}{P_1.P_3 P_2.P_4}, \qquad w = \frac{P_1.P_2 P_3.P_4}{P_1.P_4 P_2.P_3},
\end{eqnarray}
where $\sigma_i = \Delta_i + l_i$. $G^{\Delta_1,\Delta_2,\Delta_3,\Delta_4}_{{\cal O},k}(u,w)$ is the conformal block corresponding to exchange of operator $\cal{O}$ and $Q_{k}$s are tensor structure that appear due to the external spinning operators. It will be convenient to define another cross-ratio $v$, where 
\begin{equation}
v=\frac{u}{w}.
\end{equation} We will be using both the sets $(u, v)$ and $(u, w)$ throughout the paper. The spinning conformal block is fairly complicated and in practice we will need the leading two exchanges for our purpose. This was achieved in \cite{Costa:2011dw}, by applying the differential operator corresponding to the three point function of the two external operators and the exchanged operator  onto the scalar conformal block. Assuming the lowest twist operator after the identity exchange is the stress tensor, the resulting spinning conformal blocks in the embedding space formalism are then obtained from the scalar blocks in the following manner \cite{Li:2015itl, Costa:2011dw},
\begin{eqnarray}\label{spinningblocksschannel}
\langle j(P_1,Z_1)j(P_2,Z_2) \phi(P_3) \phi(P_4) \rangle &=& C_J \frac{\alpha_1 H_{12}}{P_{12}^3 P_{34}^{\Delta_\phi}} + \frac{\lambda_{\phi\phi T}}{ \sqrt{\alpha_2 C_T}} (D_{\text{even}} - D_{\text{odd}}) {\cal W}_T(\Delta, \Delta , \Delta_\phi, \Delta_\phi)\nonumber\\
&& + 
\cdots ,\nonumber\\
\end{eqnarray} 
where the first term is the identity exchange, and $D_{\text{even}}$ and $D_{\text{odd}}$ are the parity even and parity odd operators respectively corresponding to stress tensor exchange. The pre factors have been fixed by comparing the leading term in the conformal block at the limit $u\rightarrow 0, v\rightarrow 1$ with the following
\begin{eqnarray}
\langle J(0) J(\infty) \phi(z,\bar{z}) \phi(1,1)\rangle \sim \langle J(0) J(\infty) T(1,1)\rangle.
\end{eqnarray}
  
 The $s$-channel scalar block with operator ${\cal O}$, with dimensions $\Delta_{\mathcal{O}} $ and spin $l_{\mathcal{O}}$, exchanged in the embedding formalism is given by,
\begin{eqnarray}\label{scblock}
{\cal W}_{{\mathcal{O}}}(\Delta_1, \Delta_2 , \Delta_3, \Delta_4) &=& \left( \frac{P_2.P_4}{P_1.P_4} \right)^{\frac{1}{2}\Delta_{12}}  \left( \frac{P_1.P_4}{P_1.P_3} \right)^{\frac{1}{2}\Delta_{34}} \frac{G^{(\Delta_1,\Delta_2,\Delta_3,\Delta_4)}_{{\cal O}}(u,w)}{(P_1.P_2)^{\frac{1}{2}(\Delta_1+\Delta_2)}(P_3.P_4)^{\frac{1}{2}(\Delta_3+\Delta_4)}} \nonumber\\
&&\times \frac{1}{(-2)^{\frac{1}{2}(\Delta_1+\Delta_2+\Delta_3+\Delta_4)}},\end{eqnarray}
where $u = \frac{P_1.P_2 P_3.P_4}{P_1.P_3 P_2.P_4}$ and $
w = \frac{P_1.P_2 P_3.P_4}{P_1.P_4 P_2.P_3}$ and the factor $(-2)^{\frac{1}{2}(\Delta_1+\Delta_2+\Delta_3+\Delta_4)}$ has been introduced to match with the scalar block in projected coordinates.
The parity even operators are given by,

\begin{eqnarray}\label{deven}
D_{\text{even}}{\cal W}_T(\Delta, \Delta , \Delta_\phi, \Delta_\phi) &=& \alpha_3\left(\left(2 \lambda_{jjT} -\frac{3 C_J}{8 \pi }\right)D_{11}D_{22} +  \left(2 \lambda_{jjT} -\frac{9 C_J}{8 \pi }\right)D_{12}D_{21}-2 \lambda_{jjT} H_{12}\right)\nonumber\\
&&\times \sum^{1,1} {\cal W}_T(\Delta, \Delta , \Delta_\phi, \Delta_\phi),\nonumber\\ 
\end{eqnarray}
where $\lambda_{jjT}$ and $C_J$ are the two independent parameters of the three point function of $\langle JJT\rangle$. They are related to the parametrizations of \cite{Osborn:1993cr} as follows,
\begin{eqnarray}
c=\lambda_{jjT}, \qquad 2S_d(c+e)=C_J.
\end{eqnarray}
The normalization factors $\alpha_1, \alpha_2$ are fixed by matching it with the two point function in \citep{Osborn:1993cr}.
\begin{eqnarray}
\alpha_1= \alpha_2= 1
\end{eqnarray}
The three point function coupling $\lambda_{\phi\phi T}$ has been fixed as follows. From \citep{Li:2015itl} the three point function of two scalars and a conserved stress tensor in embedding space is given as follows 
\begin{eqnarray}
\langle \phi(P_1) \phi(P_2) T(Z_3,P_3) &=& \hat{\lambda}_{\phi\phi T} \frac{V_3^2}{(P_{12})^{(\Delta_\phi -1-\frac{d}{2})}(P_{13})^{(\frac{d}{2}+1)}(P_{23})^{(\frac{d}{2}+1)}}.
\end{eqnarray}
We demand that in projected coordinates, this matches with the three point function of \citep{Osborn:1993cr}. 
\begin{eqnarray}
\hat{\lambda}_{\phi\phi T} &=& -\frac{\Delta_\phi d}{(d-1)s_d},\qquad \lambda_{\phi\phi T} = -\frac{\Delta_\phi d}{(d-1)s_d \sqrt{\alpha_2 C_T}}.
\end{eqnarray}
The normalization constant $\alpha_3$ is fixed in the following manner. Starting from the three point function $\langle \phi \phi T \rangle$ (after stripping off $\hat{\lambda}_{\phi \phi T}$), we need to act the differential operator on the three point function to get the $\langle JJT \rangle$ of \cite{Osborn:1993cr} 
\begin{eqnarray}
\alpha_3 &=& 1.
\end{eqnarray} 
The parity odd differential operators are similarly given by, 
\begin{eqnarray}\label{dodd}
D_{\text{odd}}{\cal W}_T(\Delta, \Delta , \Delta_\phi, \Delta_\phi) &=&-\frac{ip_j}{1152} \left({\cal D}^{(3)} + {\cal D}^{(4)}\right){\cal W}_T(\Delta, \Delta , \Delta_\phi, \Delta_\phi),\nonumber\\
{\cal D}^{(3)} &=& 8\tilde{D}_1D_{21}\sum^{1,0}+4\tilde{D}_1D_{22}\sum^{0,1}, \nonumber\\
{\cal D}^{(4)} &=& 8\tilde{D}_2D_{12}\sum^{0,1}+4\tilde{D}_2D_{11}\sum^{1,0}.\nonumber\\
\end{eqnarray}
The explicit expressions for the differential operators are
\begin{eqnarray}\label{scdiffops}
D_{11} &=& (P_1.P_2)(Z_1.\frac{\partial}{\partial P_2})-(Z_1.P_2)(P_1.\frac{\partial}{\partial P_2})-(Z_1.Z_2)(P_1.\frac{\partial}{\partial Z_2})+(P_1.Z_2)(Z_1.\frac{\partial}{\partial Z_2}),\nonumber\\
D_{12} &=& (P_1.P_2)(Z_1.\frac{\partial}{\partial P_1})-(Z_1.P_2)(P_1.\frac{\partial}{\partial P_1})+(Z_1.P_2)(Z_1.\frac{\partial}{\partial Z_1}) ,\nonumber\\
\tilde{D}_1 &=& \epsilon \left( Z_1, P_1, \frac{\partial}{\partial P_1}, P_2, \frac{\partial}{\partial P_2} \right) + \epsilon \left( Z_1, P_1, \frac{\partial}{\partial P_1}, Z_2, \frac{\partial}{\partial Z_2} \right) ,\nonumber\\
\tilde{D}_2 &=& \epsilon \left( Z_2, P_2, \frac{\partial}{\partial P_2}, P_1, \frac{\partial}{\partial P_1} \right) + \epsilon \left( Z_2, P_2, \frac{\partial}{\partial P_2}, Z_1, \frac{\partial}{\partial Z_1} \right). \nonumber\\   
\end{eqnarray}

The other two differential operators $D_{21}$ and $D_{22}$ can be obtained from the above equations by $1 \leftrightarrow 2$. 
 The operator $\sum^{a,b}$ acts on the scalar block ${\cal W}_{\mathcal{O}}(\Delta, \Delta , \Delta_\phi, \Delta_\phi)$ as follows
\begin{eqnarray}
\sum^{a,b}{\cal W}_{\mathcal{O}}(\Delta, \Delta , \Delta_\phi, \Delta_\phi)&=& {\cal W}_{\mathcal{O}}(\Delta + a, \Delta + b, \Delta_\phi, \Delta_\phi).
\end{eqnarray}
The action of the differential operators $D_{\text{even}}$ and $D_{\text{odd}}$ are explicitly given in the Appendix \ref{lhsdiffops}.
 \paragraph{ } Let us review the implications of crossing symmetry for the parity even part of the four point function $\langle J(P_1,Z_1) J(P_2,Z_2) \phi (P_3) \phi (P_4) \rangle$. The crossing equation can be written down, in embedding space, as follows \citep{Li:2015itl} 
 \begin{eqnarray}\label{parityevencrossingeqn}
&&C_J \frac{H_{12}}{P_{12}^3 P_{34}^{\Delta_\phi}} + \frac{\lambda_{\phi\phi T}}{ \sqrt{C_T}} D_{\text{even}} {\cal W}_T(2, 2, \Delta_\phi, \Delta_\phi) \nonumber\\
&&=\sum_{\tau, l} P_{[j,\phi]_{\tau,l}} D^t_{[j,\phi]_{\tau,l}} {\cal W}_{[j,\phi]_{\tau,l}}^t(2, 2, \Delta_\phi, \Delta_\phi) + P_{\tilde{[j,\phi]}_{\tau,l}} D^t_{\tilde{[j,\phi]}_{\tau,l}} {\cal W}_{\tilde{[j,\phi]}_{\tau,l}}^t(2, 2, \Delta_\phi, \Delta_\phi)\nonumber\\
&&+\sum_{\tau, l} P_{[j,\phi]_{\tau,l}} \gamma_{[j,\phi]_{\tau,l}} \partial_\tau D^t_{[j,\phi]_{\tau,l}} {\cal W}_{[j,\phi]_{\tau,l}}^t(2, 2, \Delta_\phi, \Delta_\phi)\nonumber\\
&&+ P_{\tilde{[j,\phi]}_{\tau,l}}\gamma_{\tilde{[j,\phi]}_{\tau,l}} \partial_{\tilde{\tau}} D^t_{\tilde{[j,\phi]}_{\tau,l}} {\cal W}_{\tilde{[j,\phi]}_{\tau,l}}^t(2, 2, \Delta_\phi, \Delta_\phi).\nonumber\\
\end{eqnarray}
where ${\cal W}_{\mathcal{O}}$ is given by eqn \eqref{scblock} and action of the differential operators $D_{\text{even}}$ (eqn \eqref{deven}) is given in Appendix \ref{lhsdiffops}. The $t$-channel scalar blocks ${\cal W}_{\mathcal{O}}^t(\Delta_1, \Delta_2, \Delta_3, \Delta_4)$ are obtained from eqn \eqref{scblock} by $2 \leftrightarrow 4$. The first term on the LHS denotes the identity exchange in the $s$-channel while the second term denotes the first correction to identity. The identity exchange produces the OPE coefficients $P_{[j,\phi]_{\tau,l}}$ and $P_{\tilde{[j,\phi]}_{\tau,l}}$  on the RHS while the stress tensor exchange is responsible for the anomalous dimensions $\gamma_i$. Note that in three dimensions, we have a unique $D_{\text{odd}}  {\cal W}_T(2, 2, \Delta_\phi, \Delta_\phi)$ contribution apart from the usual parity even one which was discussed in \citep{Li:2015itl}. For now we just review the parity even case, we will discuss implications of this modification to the crossing equation in detail in section \ref{opeandanomalousdimensions}. 
The presence of the operators $D^t_{[j,\phi]_{\tau,l}}$ and $D^t_{\tilde{[j,\phi]}_{\tau,l}}$ on the RHS is due to the fact that crossing symmetry implies that the $t$-channel receives contributions from two families of double twist operators with twists given by \citep{Li:2015itl},
\begin{eqnarray}\label{parityevendoubletwistops}
[j,\phi]_{\tau,l} &=& J_\nu (\partial^{2n}) \partial_{\mu_1}\partial_{\mu_2}\cdots\partial_{\mu_{l-1}} \phi, \qquad \tau_{[j,\phi]} = \Delta_\phi +1 +2n ,\nonumber\\
\tilde{[j,\phi]}_{\tau,l} &=& \epsilon_{k\nu \rho}J^\nu \partial^\rho (\partial^{2n}) \partial_{\mu_1}\partial_{\mu_2}\cdots\partial_{\mu_{l-1}} \phi, \qquad \tau_{\tilde{[j,\phi]}} = \Delta_\phi +2 +2n. \nonumber\\
\end{eqnarray}
We restrict our analysis to $n=0$ (and therefore $\tau_0$) cases. The operators $D_{[j,\phi]_{\tau_0,l}}$ and $D_{\tilde{[j,\phi]}_{\tau_0,l}}$ can be fixed by demanding that they reproduce the correct three point function $\langle J \phi [J, \phi] \rangle$ and the action of $D^t_{[j,\phi]_{\tau_0,l}}$ and $D^t_{\tilde{[j,\phi]}_{\tau_0,l}}$on the blocks ${\cal W}_{\mathcal{O}}^t(2, 2, \Delta_\phi, \Delta_\phi)$ are given as follows,
\begin{eqnarray}\label{tcblocks}
D^t_{[j,\phi]_{\tau_0,l}} &=& \left( \frac{-1}{l+\Delta_\phi -1} D^t_{11} \sum^{1,0}_L + D^t_{12} \sum^{0,1}_L\right)\left( \frac{-1}{l+\Delta_\phi -1} D^t_{44} \sum^{1,0}_R + D^t_{43} \sum^{0,1}_R\right), \nonumber\\
D^t_{\tilde{[j,\phi]}_{\tau_0,l}} &=& \tilde{D}^t_1\tilde{D}^t_4,\nonumber\\
\end{eqnarray}
where the $t$-channel differential operators $D^t_{ij}$ and $\tilde{D}^t_i$ , denoted by the superscript $t$, are derived from equation \eqref{scdiffops} by the $(2 \leftrightarrow 4)$. 
  
\begin{eqnarray}\label{parityevenoddbuildingblocksrhs}
D^t_{11} &=& (P_1.P_4)(Z_1.\frac{\partial}{\partial P_4})-(Z_1.P_4)(P_1.\frac{\partial}{\partial P_4})-(Z_1.Z_4)(P_1.\frac{\partial}{\partial Z_4})+(P_1.Z_4)(Z_1.\frac{\partial}{\partial Z_4}),\nonumber\\
D^t_{12} &=& (P_1.P_4)(Z_1.\frac{\partial}{\partial P_1})-(Z_1.P_4)(P_1.\frac{\partial}{\partial P_1})+(Z_1.P_4)(Z_1.\frac{\partial}{\partial Z_1}) ,\nonumber\\
D^t_{44} &=& (P_2.P_3)(Z_2.\frac{\partial}{\partial P_3})-(Z_2.P_3)(P_2.\frac{\partial}{\partial P_3})-(Z_2.Z_3)(P_2.\frac{\partial}{\partial Z_3})+(P_2.Z_3)(Z_2.\frac{\partial}{\partial Z_3}),\nonumber\\
D^t_{43} &=& (P_2.P_3)(Z_2.\frac{\partial}{\partial P_2})-(Z_2.P_3)(P_2.\frac{\partial}{\partial P_2})+(Z_2.P_3)(Z_2.\frac{\partial}{\partial Z_2}), \nonumber\\
\tilde{D}^t_1 &=& \epsilon \left( Z_1, P_1, \frac{\partial}{\partial P_1}, P_4, \frac{\partial}{\partial P_4} \right) + \epsilon \left( Z_1, P_1, \frac{\partial}{\partial P_1}, Z_4, \frac{\partial}{\partial Z_4} \right), \nonumber\\
\tilde{D}^t_4 &=& \epsilon \left( Z_2, P_2, \frac{\partial}{\partial P_2}, P_3, \frac{\partial}{\partial P_3} \right) + \epsilon \left( Z_2, P_2, \frac{\partial}{\partial P_2}, Z_3, \frac{\partial}{\partial Z_3} \right) .\nonumber\\
\end{eqnarray}
The shifts $\sum^{i,j}_R$ and $\sum^{i,j}_L$ are done as follows
\begin{eqnarray}
\sum^{i,j}_R {\cal W}^t_{\mathcal{O}}(\Delta_1, \Delta_2 , \Delta_3, \Delta_4) &=& {\cal W}^t_{\mathcal{O}}(\Delta_1, \Delta_2+i , \Delta_3+j, \Delta_4), \nonumber\\
\sum^{i,j}_L {\cal W}^t_{\mathcal{O}}(\Delta_1, \Delta_2 , \Delta_3, \Delta_4) &=& {\cal W}^t_{\mathcal{O}}(\Delta_1+i, \Delta_2, \Delta_3, \Delta_4+j).
\end{eqnarray}
We study and obtain the OPE coefficients $P_{[j,\phi]_{\tau,l}}$ and $P_{\tilde{[j,\phi]}_{\tau,l}}$ and obtain the respective anomalous dimensions. We will be discarding the differential operators subleading in $l$ in our discussion. 
\subsection{Parity even and odd}
\paragraph{ } We solve the bootstrap equation in the limit $u \rightarrow 0$, $v \rightarrow 0$. By conformal symmetry, we can put our operators as follows.
\begin{eqnarray}
&&\langle \epsilon.J(0) \epsilon.J(\infty) \phi(z,\bar{z}) \phi(1,1)  \rangle, \nonumber\\
&&u = (1-z)(1-\bar{z}) ,\qquad v = z\bar{z}.
\end{eqnarray} In the limit $u\ll v$ $\;$ \footnote{Here $v=\frac{u}{w}$.}, the closed form expressions for the scalar blocks for an exchange of operator $\mathcal{O}$ of dimensions $\Delta_m$ and spin $l_m$ on the LHS is explicitly known  \cite{Dolan:2011dv},
\begin{eqnarray}
G_{\mathcal{O}}^{(\Delta_1,\Delta_2,\Delta_3,\Delta_4)}(u, v) &=& u^{\frac{1}{2}(\Delta_m-l_m)} \left(\frac{-1}{2}\left(1-v\right)\right)^{l_m} \nonumber\\
&&\, _2F_1\left(\frac{\Delta_m + l_m + \Delta_2 - \Delta_1}{2},\frac{\Delta_m + l_m + \Delta_3 - \Delta_4}{2} ;\Delta_m + l_m;1- v\right).\nonumber\\
\end{eqnarray}  
 We evaluate the parity even operators on LHS by explicitly acting on the scalar conformal block  by the differential operators $D_{\text{even}}$ and $D_{\text{odd}}$. 
 The details of this can be found in Appendix \ref{lhsdiffops}. We now solve the bootstrap equation to leading order in $v$.  This is formally solved by summing over the RHS conformal blocks corresponding to the two classes of double twist operators at large spin \cite{Fitzpatrick:2012yx, Li:2015itl}. The closed form expression for the large spin t-channel scalar block for exchange of operator of dimension $\Delta$ and spin $l$ on the RHS is given by,
 \begin{eqnarray}
 G_{\mathcal{O}'}^{(\Delta_1,\Delta_2,\Delta_3,\Delta_4)}(v, u) &=& \frac{\sqrt{l} 2^{l+\tau } v^{\tau /2} u^{\frac{\Delta_1 + \Delta_2 -\Delta_3 -\Delta_4}{4}} K_{\frac{\Delta_1 + \Delta_2 -\Delta_3 -\Delta_4}{2}}\left(2 l \sqrt{u}\right)}{\sqrt{\pi }}.
\end{eqnarray}
Note that this is in the limit $u\ll v$ in the t-channel. We assume the structure of the OPE coefficients 
 to be $\frac{A_i B_i^l}{2^{l}}$. With this ansatz the sums are performed by the following integral
 \begin{eqnarray}
\int dl l^\alpha K_\nu (2lx) &=& \frac{x^{-(\alpha+1)}}{4} \Gamma\left(\frac{1+\alpha-\nu}{2}\right)\Gamma\left(\frac{1+\alpha+\nu}{2}\right).
\end{eqnarray} 
For simplification of computation, the sums are performed first over the scalar blocks before acting on by the differential operators.
 Instead of directly acting on the parity odd differential operators in eqn \eqref{tcblocks}, we adopt the methodology used in \cite{Hofman:2016awc, Rejon-Barrera:2015bpa} to evaluate the action of parity odd differential operators on the scalar block.
\begin{eqnarray}
 D^t_{\tilde{[j,\phi]}_{\tau,l}} {\cal W}_{\tilde{[j,\phi]}_{\tau,l}}^t(2, 2, \Delta_\phi, \Delta_\phi) \sim G^{(\Delta_1,\Delta_2,\Delta_3,\Delta_4)}_{\tau, l}(v,u)\left(m^{(\mu\nu),(1,2)} + \frac{2}{\sqrt{v}}k^{(142),\mu}k^{(213),\nu} \right).\nonumber\\
\end{eqnarray}
where,
\begin{eqnarray}
m^{(\mu\nu),{i,j}} &=& \eta^{\mu\nu}-\frac{2}{x_{ij}^2}(x_{ij})^\mu(x_{ij})^\nu,\nonumber\\
k^{(ijk),\mu} &=& \frac{x_{ij}^2(x_{ik})^\mu - x_{ik}^2(x_{ij})^\mu}{(x_{ik}^2x_{ij}^2x_{jk}^2)^{\frac{1}{2}}}.
\end{eqnarray}
We obtain the following OPE coefficients and anomalous dimensions \citep{Li:2015itl}
\begin{eqnarray}
P_{[j,\phi]_{\tau_0, l}} &=& \frac{\sqrt{\pi } C_J 2^{-\Delta_\phi-l+1} l^{\Delta_\phi-\frac{1}{2}}}{\Gamma (\Delta_\phi)},\qquad P_{\tilde{[j,\phi]}_{\tau_0,l}}= \frac{\sqrt{\pi } C_J 2^{-\Delta_\phi- l-1} l^{\Delta_\phi+\frac{1}{2}}}{\Gamma (\Delta_\phi)}\nonumber\\
\gamma_{[j,\phi]_{\tau_0,l}} &=&\frac{32 \lambda_{\phi\phi T} (3 C_J-8 \pi  \lambda_{jjT} ) \Gamma (\Delta_\phi)}{3 \pi ^{5/2} C_J \sqrt{C_T} l \Gamma \left(\Delta_\phi-\frac{1}{2}\right)},\qquad \gamma_{\tilde{[j,\phi]}_{\tau,l}}= \frac{64 \lambda_{\phi\phi T} (16 \pi \lambda_{jjT} - 3 C_J) \Gamma (\Delta_\phi)}{3 \pi ^{5/2} C_J \sqrt{C_T} l \Gamma \left(\Delta_\phi-\frac{1}{2}\right)}.\nonumber\\
\end{eqnarray}
The computation was done by looking at the polarizations $(++)$ and $(xx)$. We note that the anomalous dimensions are a function of the party even OPE coefficients of $\langle JJT \rangle$. The anomalous dimensions are as a result of the matching the logarithms that one gets due to the stress tensor exchange block on the LHS with the RHS of the crossing equation. We also note that 
while only the composite operators $[j,\phi]_{\tau,l}$ contributed to the $(++)$ polarization, both $\tilde{[j,\phi]}_{\tau,l}$ and $[j,\phi]_{\tau,l}$ contributed to the RHS sum for the $(xx)$ polarization \cite{Li:2015itl}. Schematically,
\begin{eqnarray}
H^{s,++}_T(v,u) \sim H_{[j,\phi]_{\tau,l}}^{t,++}(v,u),\nonumber\\
H^{s,xx}_T(v,u) \sim H_{[j,\phi]_{\tau,l}}^{t,xx}(v,u) + H_{\tilde{[j,\phi]}_{\tau,l}}^{t,xx}(v,u).   
\end{eqnarray} 
This leads to a non trivial subtraction rule while evaluating the collider bounds from bootstrap \cite{Hofman:2016awc}. By demanding that the anomalous dimensions be negative \cite{Li:2015itl}, we obtain the parity even collider bounds in $d=3$. We have also checked that the crossing equation is satisfied for $(+-), (-+)$ and $(--)$ polarizations.

\subsection{Mixed operators}\label{opeandanomalousdimensions}
\paragraph{ }In this section we study the modifications to the crossing equations due to the parity odd stress tensor exchange in the $s$-channel. The modifications to the RHS of the crossing equation due to the presence of the parity odd term on the LHS is as follows
\begin{eqnarray}\label{crossingequation}
&& C_J \frac{ H_{12}}{P_{12}^3 P_{34}^{\Delta_\phi}} + \frac{\lambda_{\phi\phi T}}{ \sqrt{C_T}} (D_{\text{even}} - D_{\text{odd}}) {\cal W}_T(2,2 , \Delta_\phi, \Delta_\phi)\nonumber\\
&&=\sum_{\tau, l} P_{[j,\phi]_{\tau,l}} D^t_{[j,\phi]_{\tau,l}} {\cal W}^t(2, 2, \Delta_\phi, \Delta_\phi) + P_{\tilde{[j,\phi]}_{\tau,l}} D^t_{\tilde{[j,\phi]}_{\tau,l}} {\cal W}^t(2, 2, \Delta_\phi, \Delta_\phi)\nonumber\\
&&+\sum_{\tau, l} P_{[j,\phi]_{\tau,l}} \gamma_{[j,\phi]_{\tau,l}} \partial_\tau D^t_{[j,\phi]_{\tau,l}} {\cal W}^t(2, 2, \Delta_\phi, \Delta_\phi)+ P_{\tilde{[j,\phi]}_{\tau,l}}\gamma_{\tilde{[j,\phi]}_{\tau,l}} \partial_{\tilde{\tau}} D^t_{\tilde{[j,\phi]}_{\tau,l}} {\cal W}^t(2, 2, \Delta_\phi, \Delta_\phi)\nonumber\\
&&+ P^{11}_{m,\tau,l} D^{11}_m {\cal W}_{\tilde{\mathcal{O}}}^t(2, 2, \Delta_\phi, \Delta_\phi) +P^{12}_{m,\tau,l} D^{12}_m {\cal W}_{\tilde{\mathcal{O}}}^t(2, 2, \Delta_\phi, \Delta_\phi) \nonumber\\
&&+ P^{21}_{m,\tau,l} D^{21}_m {\cal W}_{\tilde{\mathcal{O}}}^t(2, 2, \Delta_\phi, \Delta_\phi) +P^{22}_{m,\tau,l} D^{22}_m {\cal W}_{\tilde{\mathcal{O}}}^t(2, 2, \Delta_\phi, \Delta_\phi), \nonumber\\
\end{eqnarray}
where $\tilde{\mathcal{O}}$ are spin $l$ double twist operators with twist given by
\begin{eqnarray}
\tau_{\tilde{\mathcal{O}}} = 1+\Delta_\phi + 2n.  
\end{eqnarray}
The operators $D^{ij}_m $ are modifications to the crossing equations that have not previously been studied. The differential operators are given to be 
\begin{eqnarray}\label{tcblocksmixed}
D_m^{11}&=& \tilde{D}^t_1 D^t_{43} \sum^{0,1}_R, \qquad D_m^{12}= \tilde{D}^t_1 D^t_{44} \sum^{1,0}_R \nonumber\\ 
D_m^{21}&=& \tilde{D}^t_4 D^t_{12} \sum^{0,1}_L,\qquad D_m^{22} = \tilde{D}^t_4 D^t_{11} \sum^{1,0}_L \nonumber\\ 
\end{eqnarray}
Their action on the scalar block is given in Appendix \ref{evenrhsops}. $P^i_{m,\tau,l}$s are the OPE coefficients corresponding to the respective differential operators. As we shall see their presence is essential for the crossing equation to be satisfied in presence of the parity violating terms in eqn \eqref{crossingequation}. 
In order to evaluate the contribution of the parity odd operators in the s-channel to the crossing equation, we look at the polarizations $(+x),(x+),(-x)$ and $(x-)$. For this choice of polarizations only the tensor structures corresponding to $ P^i_{m,\tau,l}$ s are non-zero and on the LHS, $D_{\text{even}} {\cal W}_T(2, 2, \Delta_\phi, \Delta_\phi) $ also drop out. Thus the inclusion of these extra structures on the RHS is essential for the crossing equation to be satisfied. Moreover we have also checked that the LHS parity odd conformal block do not contribute to the corrections to the OPE coefficients $P_{[j,\phi]_{\tau,l}}$ and $P_{\tilde{[j,\phi]}_{\tau,l}}$. The crossing equation then becomes
\begin{eqnarray}\label{mixedcrossing}
&&\frac{-\lambda_{\phi\phi T}}{\sqrt{C_T}} D_{\text{odd}} {\cal W}(2, 2, \Delta_\phi, \Delta_\phi) =\nonumber\\ 
&&\sum_{\tau, l} P^{11}_{m,\tau,l} D^{11}_m {\cal W}_{\tilde{\mathcal{O}}}^t(2, 2, \Delta_\phi, \Delta_\phi) +P^{12}_{m,\tau,l} D^{12}_m {\cal W}_{\tilde{\mathcal{O}}}^t(2, 2, \Delta_\phi, \Delta_\phi) \nonumber\\
&&+ P^{21}_{m,\tau,l} D^{21}_m {\cal W}_{\tilde{\mathcal{O}}}^t(2, 2, \Delta_\phi, \Delta_\phi) +P^{22}_{m,\tau,l} D^{22}_m {\cal W}_{\tilde{\mathcal{O}}}^t(2, 2, \Delta_\phi, \Delta_\phi). \nonumber\\
\end{eqnarray} 
where  the operators $D^{ij}_m$s  are given in eqn \eqref{tcblocksmixed}           , while the action of the $s$-channel $D_{\textrm{odd}}$ operators is given in 
Appendix \ref{parityodddifflhs2}. Note that the differential operators $D_{\text{odd}}$ are fundamentally different from their parity even counterpart. Because of the asymmetric shifts in the differential operators, the light cone block will therefore not have any logarithm term corresponding to $D_{\text{odd}}$ operator. To be explicit, let us look at the shifted blocks for both the parity even and parity odd differential operator for stress tensor exchange, before the action of the differential operator itself. In $d=3$ the shifted block corresponding to the stress tensor exchange in the parity even sector ($u\ll v$) is given by
\begin{eqnarray}
\sum^{1,1}G_T^{(2,2,\Delta_\phi,\Delta_\phi)}(u,v)&=&G_T^{(3,3,\Delta_\phi,\Delta_\phi)}(u,v)\nonumber\\
&=&\frac{1}{4} \sqrt{u} (v-1)^2 \, _2F_1\left(\frac{5}{2},\frac{5}{2};5;1-v\right),\nonumber\\
\end{eqnarray}    
as $ u \rightarrow 0, v \rightarrow 0$, this develops a logarithmic singularity. In contrast, for the parity odd blocks we have closed form polynomial expressions for the shifted blocks for $u\ll v$.
\begin{eqnarray}
\sum^{0,1}G_T^{(2,2,\Delta_\phi,\Delta_\phi)}(u,v)&=&G_T^{(2,3,\Delta_\phi,\Delta_\phi)}(u,v) \nonumber\\
&=&\frac{4 \sqrt{u} (1-v)^2}{\left(\sqrt{v}+1\right)^4 \sqrt{v}}.
\end{eqnarray} 
This does not have a logarithmic singularity as $u \rightarrow 0, v \rightarrow 0$.
 Proceeding similarly as before, with the epsilon tensors evaluated with the help of Appendix \ref{epsilontensors} and the following ansatz for the OPE coefficients, 
\begin{eqnarray}
P^{11}_{m,\tau,l} &=& \frac{B l^A}{2^{l}},\qquad P^{12}_{m,\tau,l} = \frac{B_2 l^{A_2}}{2^{l}}\nonumber\\
P^{21}_{m,\tau,l} &=& \frac{Y l^X}{2^{l}},\qquad P^{22}_{m,\tau,l} = \frac{Y_2 l^{X_2}}{2^{l}}
\end{eqnarray}
we have for the $(+x)$ polarization ,
\begin{eqnarray}
&&\frac{2p_j u^{\frac{1}{2}-\Delta_\phi}\lambda_{\phi\phi T}}{\sqrt{C_T}}\nonumber\\
&& =\frac{ -i B  2^{-1+\tau } \Gamma \left(\frac{1}{2} (A-\Delta_\phi+5)\right) \Gamma \left(\frac{1}{2} (A+\Delta_\phi+2)\right) u^{\frac{1}{2} (-A-\Delta_\phi)} v^{\frac{\tau -\Delta_\phi}{2}}}{\sqrt{\pi } }\nonumber\\
&&+\frac{i B_2  2^{-1+\tau} \Gamma \left(\frac{1}{2} (A_2-\Delta_\phi+6)\right) \Gamma \left(\frac{1}{2} (A_2+\Delta_\phi+1)\right) u^{\frac{1}{2} (-A_2-\Delta_\phi-1)} v^{\frac{\tau -\Delta_\phi}{2}}}{\sqrt{\pi } } \nonumber\\
&&-i\frac{  Y 2^{-1+\tau } u^{\frac{1}{2} (-\Delta_\phi-X-2)} v^{\frac{1}{2} (-\Delta_\phi+\tau -1)} \Gamma \left(\frac{1}{2} (X-\Delta_\phi+7)\right) \Gamma \left(\frac{1}{2} (X+\Delta_\phi+2)\right)}{\sqrt{\pi } }\nonumber\\
&&-i\frac{  Y_2 2^{-2+\tau} (\Delta_\phi+\tau -3) u^{\frac{1}{2} (-\Delta_\phi-X_2-1)} v^{\frac{1}{2} (-\Delta_\phi+\tau -1)} \Gamma \left(\frac{1}{2} (X_2-\Delta_\phi+6)\right) \Gamma \left(\frac{1}{2} (X_2+\Delta_\phi+1)\right)}{\sqrt{\pi } }.\nonumber\\
\end{eqnarray}
Similarly for the $(x+)$ polarization we have
\begin{eqnarray}
&&\frac{-2 p_j u^{\frac{1}{2}-\Delta_\phi}\lambda_{\phi\phi T}}{\sqrt{C_T}\sqrt{v}}\nonumber\\
&&= \frac{-iB (v-1) 2^{-1+\tau} \Gamma \left(\frac{1}{2} (A-\Delta_\phi+7)\right) \Gamma \left(\frac{1}{2} (A+\Delta_\phi+2)\right) u^{\frac{1}{2} (-A-\Delta_\phi-2)} v^{\frac{1}{2} (-\Delta_\phi+\tau -2)}}{\sqrt{\pi } }\nonumber\\
&&+i\frac{ B_2  2^{-2+\tau} \Gamma \left(\frac{1}{2} (A_2-\Delta_\phi+6)\right) \Gamma \left(\frac{1}{2} (A_2+\Delta_\phi+1)\right) u^{\frac{1}{2} (-A_2-\Delta_\phi-1)} v^{\frac{1}{2} (-\Delta_\phi+\tau -2)} (\Delta_\phi+\tau -3)}{\sqrt{\pi } }\nonumber\\
&&+i\frac{ Y 2^{-1+\tau} u^{\frac{1}{2} (-\Delta_\phi-X)} v^{\frac{1}{2} (-\Delta_\phi+\tau -1)} \Gamma \left(\frac{1}{2} (X-\Delta_\phi+5)\right) \Gamma \left(\frac{1}{2} (X+\Delta_\phi+2)\right)}{\sqrt{\pi } }\nonumber\\
&&-i\frac{ Y_2 2^{-1+\tau } u^{\frac{1}{2} (-\Delta_\phi-X_2-1)} v^{\frac{1}{2} (-\Delta_\phi+\tau -1)} \Gamma \left(\frac{1}{2} (X_2-\Delta_\phi+6)\right) \Gamma \left(\frac{1}{2} (X_2+\Delta_\phi+1)\right)}{\sqrt{\pi }}.\nonumber\\
\end{eqnarray}
For the $(-x)$ polarization,
\begin{eqnarray}
&&-\frac{2p_j u^{\frac{1}{2}-\Delta_\phi}\lambda_{\phi\phi T}}{\sqrt{C_T}\sqrt{v}}\nonumber\\
&&=\frac{ -i(-1+v) B  2^{-1+\tau} \Gamma \left(\frac{1}{2} (A-\Delta_\phi+5)\right) \Gamma \left(\frac{1}{2} (A+\Delta_\phi+2)\right) u^{\frac{1}{2} (-A-\Delta_\phi-2)} v^{\frac{1}{2} (-\Delta_\phi+\tau -2)}}{\sqrt{\pi } }\nonumber\\
&&-i\frac{ B_2 2^{-1+\tau} \Gamma \left(\frac{1}{2} (A_2-\Delta_\phi+6)\right) \Gamma \left(\frac{1}{2} (A_2+\Delta_\phi+1)\right) u^{\frac{1}{2} (-A_2-\Delta_\phi-1)} v^{\frac{1}{2} (-\Delta_\phi+\tau -2)}}{\sqrt{\pi } }\nonumber\\
&&+i\frac{ Y 2^{-2+\tau} (\Delta_\phi-\tau -1) u^{\frac{1}{2} (-\Delta_\phi-X-2)} v^{\frac{1}{2} (-\Delta_\phi+\tau -3)} \Gamma \left(\frac{1}{2} (X-\Delta_\phi+5)\right) \Gamma \left(\frac{1}{2} (X+\Delta_\phi+2)\right)}{\sqrt{\pi } }\nonumber\\
&&+i\frac{ (2 v-1) Y_2 2^{-1+\tau} u^{\frac{1}{2} (-\Delta_\phi-X_2-3)} v^{\frac{1}{2} (-\Delta_\phi+\tau -3)} \Gamma \left(\frac{1}{2} (X_2-\Delta_\phi+6)\right) \Gamma \left(\frac{1}{2} (X_2+\Delta_\phi+3)\right)}{\sqrt{\pi } }.\nonumber\\
\end{eqnarray}
For the $(x-)$ polarization,
\begin{eqnarray}
&&\frac{2 p_j u^{\frac{1}{2}-\Delta_\phi}\lambda_{\phi\phi T}}{\sqrt{C_T} }\nonumber\\
&&=i\frac{ B  2^{-2+\tau} (-\Delta_\phi+\tau +1) \Gamma \left(\frac{1}{2} (A-\Delta_\phi+5)\right) \Gamma \left(\frac{1}{2} (A+\Delta_\phi+2)\right) u^{\frac{1}{2} (-A-\Delta_\phi-2)} v^{\frac{1}{2} (-\Delta_\phi+\tau -2)}}{\sqrt{\pi } }\nonumber\\
&&+i\frac{ B_2 2^{-1+\tau} \Gamma \left(\frac{1}{2} (A_2-\Delta_\phi+6)\right) \Gamma \left(\frac{1}{2} (A_2+\Delta_\phi+3)\right) u^{\frac{1}{2} (-A_2-\Delta_\phi-3)} v^{\frac{1}{2} (-\Delta_\phi+\tau -2)} }{\sqrt{\pi } }\nonumber\\
&&-i\frac{  Y 2^{-1+\tau } u^{\frac{1}{2} (-\Delta_\phi-X-2)} v^{\frac{1}{2} (-\Delta_\phi+\tau -1)} \Gamma \left(\frac{1}{2} (X-\Delta_\phi+5)\right) \Gamma \left(\frac{1}{2} (X+\Delta_\phi+2)\right)}{\sqrt{\pi } }\nonumber\\
&&+i\frac{  Y_2 2^{-1+\tau } u^{\frac{1}{2} (-\Delta_\phi-X_2-1)} v^{\frac{1}{2} (-\Delta_\phi+\tau -1)} \Gamma \left(\frac{1}{2} (X_2-\Delta_\phi+6)\right) \Gamma \left(\frac{1}{2} (X_2+\Delta_\phi+1)\right)}{\sqrt{\pi } }.
\end{eqnarray}

A possible set of solutions is obtained for these set of equations corresponding to $\tau_0=\Delta_\phi+1$,

\begin{eqnarray}\label{mixedope}
P^{11}_{m,\tau,l} &=& -P^{21}_{m,\tau,l}=-\frac{\sqrt{\pi } 2^{-\Delta_\phi+1} ip_j \lambda_{\phi\phi T}}{\Gamma \left(\Delta_\phi-\frac{1}{2}\right)}\frac{l^{(\Delta_\phi-3)}}{\sqrt{C_T}2^{l}} ,\nonumber\\
P^{12}_{m,\tau,l} &=& -P^{22}_{m,\tau,l}=\frac{\sqrt{\pi } 2^{-\Delta_\phi+1} ip_j\lambda_{\phi\phi T}}{\Gamma \left(\Delta_\phi-\frac{1}{2}\right)}\frac{l^{(\Delta_\phi-4)}}{\sqrt{C_T} 2^{l}}.
\end{eqnarray}
The parity odd contribution on the LHS of the crossing equation is fundamentally different from its parity even counterpart. While as seen from the crossing equation eqn \eqref{parityevencrossingeqn} and discussions henceforth, the parity even part of the stress tensor exchange block contributes at one order lower than the identity exchange. It is directly responsible for the anomalous dimensions of the operators in eqn  \eqref{parityevendoubletwistops}. On the other hand the parity odd terms are responsible for a new set of OPE coefficients corresponding to the tower of operators $\tilde{\mathcal{O}}$ ($\tau= 1+\Delta_\phi + 2n$) in the $t$-channel exchange. The essential difference is due to the fact that there are no logarithm terms 
at this order in the $u \rightarrow 0$, $v \rightarrow 0$ expansion of the parity odd conformal blocks corresponding to the stress tensor exchange. 
A simple reason for the absence of logarithmic terms in the parity odd block  is that it can be 
considered as a transition between a  double trace trace operators of opposite parity in the $t$ channel. 
This is an off diagonal term and logarithms occur only in the diagonal term on expanding $v^\tau$
\footnote{We thank David Simmons-Duffins for this argument.}.

\section{Positivity constraints}\label{positivityconstraints}
                     \paragraph{ }  In this section we will study 
the implications of reflection positivity and crossing symmetry on the euclidean four point function. Following \citep{Li:2015itl, Hartman:2015lfa}, we define in light cone coordinates, 
\begin{eqnarray}
G(z,\bar{z}) &=& \langle \epsilon.J(0) \phi(z,\bar{z}) \phi(1,1) \epsilon.J(\infty) \rangle ,
\end{eqnarray}
where, $\epsilon$ denotes the polarization tensor associated with the currents. The $t$-channel stress tensor exchange for this correlator, we can get from eqns \eqref{spinningblocksschannel}, \eqref{scblock}, \eqref{deven} and \eqref{dodd} by the exchange of $2 \leftrightarrow 4$. The operators $O_1, O_3$ and $O_4$ have been inserted at a space like separation $\tau=0$, while the operator $O_2$ is taken at an arbitrary euclidean time.
\begin{eqnarray}
x_1&=&(0,0,0) \qquad x_2=(\tau,y_2,0,0), \qquad x_3 = (0,1,0), \qquad x_4= \lim_{a \rightarrow \infty}(0,a,0).\nonumber\\   
\end{eqnarray}  
Under the exchange $2 \leftrightarrow 4$, using the projection formulae eqn \eqref{projection}, the cross ratios now become,
\begin{eqnarray}
\tilde{u} &=& \frac{x_{14}^2x_{32}^2}{x_{13}^2x_{24}^2} ,\qquad \tilde{w} = \frac{x_{14}^2x_{32}^2}{x_{12}^2x_{43}^2}.
\end{eqnarray}
The light cone coordinates are defined as $z=y+i\tau$ and $\bar{z}=y-i\tau$. With the chosen kinematics, the cross-ratios become,
\begin{eqnarray}
\tilde{u} &=& (1-z)(1-\bar{z}), \qquad \tilde{w}= \frac{(1-z)(1-\bar{z})}{z\bar{z}}. 
\end{eqnarray}
In this analysis we will be interested in the limit $\bar{z} \rightarrow 1$ while $z$ is held fixed. Note that this limit is different from the limit used in the analysis of the crossing equation in the previous section. Let us define 
\begin{eqnarray}\label{coordparametrize}
z=1+\sigma, \qquad
\bar{z}=1+\eta\sigma.
\end{eqnarray}  
where $\sigma$ is complex with $\text{Im}(\sigma)\geq 0$ and $|\sigma| \leq R$, while $\eta$ is real and satisfies $0 < \eta \ll R \ll 1$. Excluding the origin, this is a small disc in the complex $\sigma$ plane and we refer to this as the region D. The light cone limit translates to $\eta \rightarrow 0$ with $\sigma$ held fixed \cite{Hartman:2015lfa}. We define the following normalized four point function following \citep{Li:2015itl},
\begin{eqnarray}\label{jphiphijcor}
G^{\mu\nu}_\eta(\sigma)&=& \frac{\langle J^\mu(0)\phi(z,\bar{z}) \phi(1,1) J^\nu(\infty))  \rangle}{\langle \phi(z,\bar{z}) \phi(1,1) \rangle},\nonumber\\
\hat{G}^{\mu\nu}_\eta(\sigma)&=& \frac{\langle J^\mu(0)\phi(ze^{-2\pi i},\bar{z}) \phi(1,1) J^\nu(\infty))  \rangle}{\langle \phi(z,\bar{z}) \phi(1,1) \rangle}.\nonumber\\
\end{eqnarray}
Following \citep{Hartman:2015lfa, Li:2015itl}, we summarise the arguments for the bounds in the following manner. The Euclidean correlator, eqn  \eqref{jphiphijcor}, has convergent expansions in the various OPE channels ($s$, $t$ and $u$-channels).  The
crux of the argument lies in the fact that, using reflection positivity in $s$-channel and $u$-channel expansions, it can be shown that the correlator $\hat{G}^{\mu\nu}_\eta(\sigma)$ is analytic in the region defined as D as well as  
\begin{equation}
\textrm{Re}\left(G^{\mu\nu}_\eta(\sigma)-\hat{G}^{\mu\nu}_\eta(\sigma)\right) \geq 0, \qquad \sigma\in[-R, R]. \nonumber
\end{equation}  
Using these two conditions we obtain a bound on the OPE coefficients appearing in the $t$-channel expansion of eqn \eqref{jphiphijcor} in the light cone limit. We proceed by evaluating the leading term in the $t$-channel stress tensor exchange of $\hat{G}^{\mu\nu}_\eta(\sigma)$ in the light cone limit. The light cone limit of the blocks $G(u,w)$ for operator ${\cal O}(\Delta_m,l_m)$ exchange are defined as \citep{Dolan:2011dv},
\begin{eqnarray}
G^{(\Delta_1,\Delta_2,\Delta_3,\Delta_4)}(\tilde{u}, \tilde{w}) &=& \tilde{u}^{\frac{1}{2}(\Delta_m-l_m)} \left(\frac{-1}{2}\left(1-\frac{\tilde{u}}{\tilde{w}}\right)\right)^{l_m} \nonumber\\
&&\, _2F_1\left(\frac{\Delta_m + l_m + \Delta_4 - \Delta_1}{2},\frac{\Delta_m + l_m + \Delta_3 - \Delta_2}{2} ;\Delta_m + l_m;1- \frac{\tilde{u}}{\tilde{w}}\right),\nonumber\\
\end{eqnarray}  
where, $\Delta_i$s indicate the dimension of the scalar operators at position $P_i$s after the shifts due to the differential operators. 
For our purposes, we consider stress tensor exchange of $\Delta_m=3$, $l_m=2$. 
\subsection{Parity even}
\paragraph{ } In this section we present the analysis for the parity even collider bounds. We look at the parity even part of the differential operator eqn \eqref{deven} (to be precise with $2 \leftrightarrow 4$). For going round the cut we apply the following analytic continuation formula
\begin{eqnarray}
\lim_{\epsilon \rightarrow 0^+}\, _2F_1(a,b;c;x + i\epsilon)&=&e^{2 i \pi  (a+b-c)} \, _2F_1(a,b;c;x)\nonumber\\
&&+\frac{2 i \pi  \Gamma (c) e^{i \pi  (a+b-c)} \, _2F_1(a,b;a+b-c+1;1-x)}{\Gamma (c-a) \Gamma (c-b) \Gamma (a+b-c+1)}\text{/;}x>1 \nonumber\\
\end{eqnarray}

We have checked that applying the differential operators before going round the cut is identical to applying the differential operators on the conformal blocks, after going round the cut. Technically the latter is simpler and we look at the polarizations $G^{++}$ and $G^{xx}$, for which there is no contribution from the parity odd differential operators eqn  \eqref{dodd}.  On the first sheet, $G^{++}$ and $G^{xx}$ have no poles as $\eta
  \rightarrow 0$ with $\sigma$ held fixed. On the second sheet,  
the correlator $\hat{G}^{\mu\nu}_\eta(\sigma)$ is  
\begin{eqnarray}\label{parityevennosubg}
\hat{G}^{++}_\eta(\sigma)&=& \lim_{a \rightarrow \infty} -\frac{2 C_J}{a^4}+\frac{256 i  \sqrt{\eta } \lambda_{\phi\phi T} (8 \pi  \lambda_{jjT} -3 C_J)}{3 \pi ^2 a^4 \sqrt{C_T} \sigma },\nonumber\\
\hat{G}^{xx}_\eta(\sigma)&=& \lim_{a \rightarrow \infty} \frac{C_J}{a^4}-\frac{256 i  \sqrt{\eta } \lambda_{\phi\phi T} (C_J-8 \pi  \lambda_{jjT} )}{3 \pi ^2 a^4 \sqrt{C_T} \sigma },\nonumber\\
\end{eqnarray} 
where
$\lambda_{\phi\phi T}=-\frac{\Delta_\phi d}{(d-1)s_d\sqrt{ C_T}}$.
 
We see that the correlators develop singularities on the second sheet after going round the branch cut.     
The above expressions for $t$-channel light cone singularity do not lead to the most optimal bounds. As illustrated using crossing symmetry in Section \ref{jjphiphifromphi}, a subtraction is needed and the reason for it is the following, if we consider crossing symmetry of the four point function $\langle JJ \phi \phi \rangle$, we find that in the dual channel, the large $l$ limit is dominated by two classes of composite operators indicated by two different twists given by eqn \eqref{parityevendoubletwistops} \footnote{Equivalently, the stress tensor exchange in the t-channel of $\langle J \phi \phi J \rangle$ correlator is reproduced by an infinite tower of  composite operators in s-channel \cite{Hofman:2016awc}.} \citep{Li:2015itl}. For the polarization $(++)$, the s-channel stress tensor exchange is reproduced in the dual channel by just the symmetric composite operator of twist $\tau=1+\Delta_\phi$ but for the polarization $(xx)$, both the classes of twists contribute. Following \citep{Hofman:2016awc}, in order to isolate the contribution of the antisymmetric part in $\hat{G}^{xx}$, we subtract in the following manner. The subtraction for the leading divergent part is $\frac{-1}{4}$ times $\hat{G}^{++}$
and the subleading contribution is $\frac{-1}{3}$ times that of $\hat{G}^{++}$. This subtraction scheme has been discussed in detail in \citep{Hofman:2016awc}\footnote{See section 4.1 and 4.4 of \cite{Hofman:2016awc}}. 
One aspect of this subtraction is that the  it  was only the relative normalisation of the $\hat{G}^{++}$
in comparison with  $\hat{G}^{xx}$ that   \citep{Hofman:2016awc} was careful about, since the over all 
normalisation was not crucial. 
However to get the bounds including the parity odd we should also be careful of the overall normalization 
of the subtraction. We  will see that $\hat{G}^{xx}$ and  $\hat{G}^{++}$ can be thought of as 
inner products in radial quantization. Therefore the normalization on performing the subtraction is given by  
\begin{eqnarray}\label{prescription}
\hat{\tilde{G}}^{xx}_\eta(\sigma)=\left(\frac{\left(\hat{G}^{xx}_{\eta, \text{leading}}(\sigma)+ \frac{1}{4} \hat{G}^{++}_{\eta, \text{leading}}(\sigma)\right)}{( 1+ \frac{1}{4}) } + \frac{\left(\hat{G}^{xx}_{\eta, \text{sub-leading}}(\sigma)+ \frac{1}{3} \hat{G}^{++}_{\eta, \text{sub-leading}}(\sigma)\right)}{(1+ \frac{1}{3} )} \right). \nonumber\\
\end{eqnarray}
Essentially we have divided each of the subtraction by the sum of the coefficients linear combination considered. 
This is what would one would expect to do if one thinks of $\hat{G}^{xx}$ and  $\hat{G}^{++}$ as norms (see section \ref{colliderbounds} and in particular discussions around eqn \eqref{statenorm}). 
We will see that this normalisation prescription as well the subtraction procedure reproduces the 
conformal collider bounds including the parity odd term as obtained using the average null energy condition. 
%
%
Performing this subtraction together with the normalisation we obtain 
\begin{eqnarray}
\hat{G}^{++}_\eta(\sigma)&=&\lim_{a \rightarrow \infty}  -\frac{2 C_J}{a^4}+\frac{256 i  \sqrt{\eta } \lambda_{\phi\phi T} (8 \pi  \lambda_{jjT} -3 C_J)}{3 \pi ^2 a^4 \sqrt{C_T} \sigma },\nonumber\\
\hat{\tilde{G}}^{xx}_\eta(\sigma)&=&\lim_{a \rightarrow \infty}\frac{2 C_J}{5a^4}+\frac{128 i \sqrt{\eta } \lambda_{\phi\phi T} (16 \pi  \lambda_{jjT} -3 C_J)}{3 a^4\pi ^2 \sqrt{C_T} \sigma }.\nonumber\\
\end{eqnarray}
In terms of the collider parameter introduced in,  \citep{Hofman:2008ar}
\begin{eqnarray}
\lambda_{jjT}=-\frac{C_J (d-2) d \pi ^{-\frac{d}{2}} \left(a_2-d^2+d\right) \Gamma \left(\frac{d}{2}\right)}{4 (d-1)^3},
\end{eqnarray}
the contributions become
\begin{eqnarray}\label{parityeveng}
\hat{G}^{++}_\eta(\sigma)&=&\lim_{a \rightarrow \infty} \frac{-2 C_J}{a^4}- \frac{32 C_J i (a_2+2)  \sqrt{\eta } \lambda_{\phi\phi T}}{a^4\pi ^2 \sqrt{C_T} \sigma },\nonumber\\
\hat{\tilde{G}}^{xx}_\eta(\sigma)&=&\lim_{a \rightarrow \infty} \frac{2 C_J}{5a^4} -\frac{32C_Ji (a_2-2)  \sqrt{\eta } \lambda_{\phi\phi T}}{ a^4\pi ^2 \sqrt{C_T} \sigma }.
\end{eqnarray}
Note that the normalization of the sub-leading term of $(xx)$ polarization matches with the $(++)$ one. 
The subtraction together with the normalization in (\ref{prescription}) ensures this. 
\subsection{Parity odd}
                      \paragraph{}      In this section we study the contribution of the parity odd terms to collider bounds. We derive the singular contribution of the parity odd conformal blocks in the light cone limit. We note that the light cone limit of the shifted stress tensor exchange conformal block appearing in eqn \eqref{deven}, are completely different from the ones in eqn \eqref{dodd}. The difference arises due to the asymmetric shift operators in the two different cases. Let us  illustrate with an example. Let us consider the $t$-channel light cone block for stress tensor exchange for a four point function of scalars in $d=3$ 
                      The analysis is carried out by taking the limit $\bar{z} \rightarrow 1$ first. 
\begin{eqnarray}
\lim_{\bar{z}\rightarrow 1}G(\tilde{u},\tilde{w})=  \frac{1}{4} \sqrt{(z-1) (\bar{z}-1)} (z -1)^2 \, _2F_1\left(\frac{5}{2},\frac{5}{2};5;1-z\right).
%
\end{eqnarray}    
As $z \rightarrow z e^{-2i\pi }$, this develops a singularity in the second sheet due to the logarithmic branch 
cut in the hypergeometric function. 
 In contrast, for the parity odd blocks we have closed form polynomial expressions for the shifted blocks in the light cone limit.
\begin{eqnarray}
G^{(3,\Delta_\phi,\Delta_\phi,2)}(\tilde{u},\tilde{w}) &=& \frac{1}{4}\sqrt{\tilde{u}} \left(1-\frac{\tilde{u}}{\tilde{w}}\right)^2 \, _2F_1\left(2,\frac{5}{2};5;1-\frac{\tilde{u}}{\tilde{w}}\right) \nonumber\\
&=& \frac{4 \sqrt{\tilde{u}} \left(\frac{\sqrt{\tilde{u}}}{\sqrt{\tilde{w}}}-1\right)^2}{\left(\frac{\sqrt{\tilde{u}}}{\sqrt{\tilde{w}}}+1\right)^2},\nonumber\\
G^{(2,\Delta_\phi,\Delta_\phi,3)}(\tilde{u},\tilde{w}) &=&\frac{4 \sqrt{\tilde{u}} \left(\frac{\sqrt{\tilde{u}}}{\sqrt{\tilde{w}}}-1\right)^2}{\left(\frac{\sqrt{\tilde{u}}}{\sqrt{\tilde{w}}}+1\right)^4 \frac{\sqrt{\tilde{u}}}{\sqrt{\tilde{w}}}}.
\end{eqnarray} 
For the $\langle j^x \phi \phi j^+ \rangle$ correlator, acting on by the differential operators we have (by applying $2 \leftrightarrow 4$ to the structures in Appendix \ref{parityodddifflhs2} and Appendix \ref{mixedrhsops}), before going round the branch cut,
\begin{eqnarray}
-\frac{ip_jI^{\text{odd}}_1}{1152} &=&( \frac{1}{9a^4 \sqrt{z \bar{z}} (z \bar{z}-1)^4}  p_j (\bar{z}-1) \bar{z}\nonumber\\
&&\sqrt{(z-1) (\bar{z}-1)} \sqrt{\frac{1}{z \bar{z}}} \left(z \left(z^3 \bar{z}^2 \left(-2 \sqrt{z \bar{z}}+\bar{z}+12\right)+z^2 \bar{z} \left(\bar{z} \left(-4 \sqrt{z \bar{z}}-4 \bar{z}+5\right)\right.\right.\right.\nonumber\\
&&\left.\left.\left.-28 \sqrt{z \bar{z}}+32\right) +z \left(\bar{z} \left(2 \bar{z} \left(9 \sqrt{z\bar{z}}-16\right)-5\right)-18 \sqrt{z \bar{z}}+4\right)+4 \bar{z} \left(7 \sqrt{z \bar{z}}-3\right)\right.\right.
\nonumber\\
&&\left.\left. +4 \sqrt{z \bar{z}}-1\right)+2 \sqrt{z \bar{z}}\right)),\nonumber\\
-\frac{ip_jI^{\text{odd}}_2}{1152} &=&\frac{1}{18 a^4 \sqrt{z \bar{z}} (z \bar{z}-1)^4} p_j (z-1) \sqrt{(z-1) (\bar{z}-1)} \left(\bar{z} \left(z^3 \bar{z}^2 (\bar{z}+4)\right.\right.\nonumber\\
&&\left.\left.+z^2 \bar{z} \left(\bar{z} \left(2 \bar{z} \left(\sqrt{z \bar{z}}-6\right)-4 \sqrt{z \bar{z}}+5\right)-18 \sqrt{z \bar{z}}+32\right)+z \left(-28 \sqrt{z \bar{z}}\right.\right.\right.\nonumber\\
&&\left.\left.\left. +\bar{z} \left(4 \bar{z} \left(7 \sqrt{z \bar{z}}-8\right)-5\right)+12\right)+4 \sqrt{z \bar{z}}+2 \bar{z} \left(9 \sqrt{z \bar{z}}-2\right)-1\right)-2 \sqrt{z \bar{z}}\right), \nonumber\\
-\frac{ip_jI^{\text{odd}}_3}{1152} &=&(\frac{1}{9 a^4 \sqrt{z \bar{z}} (z \bar{z}-1)^3}  p_j\sqrt{(z-1) (\bar{z}-1)} \left(\bar{z} \left(z \left(z \bar{z}^2 \left(-2 \sqrt{z \bar{z}}\right.\right.\right.\right.\nonumber\\
&&\left.\left.\left.\left. +z+7\right)-2 \bar{z} \left(z \sqrt{z \bar{z}}+4 \sqrt{z \bar{z}}+z-1\right)+8 \sqrt{z \bar{z}}-7\right)+2 \sqrt{z \bar{z}}-1\right)+2 \sqrt{z \bar{z}}\right)),\nonumber\\
-\frac{ip_jI^{\text{odd}}_4}{1152} &=&(\frac{1}{18 a^4 \sqrt{z \bar{z}} (z \bar{z}-1)^3} p_j \bar{z} \sqrt{(z-1) (\bar{z}-1)} \sqrt{\frac{1}{z \bar{z}}} \left(z \left(\bar{z} \left(z^2 \bar{z} \right.\right.\right.\nonumber\\
&&\left.\left.\left. \left(2 \sqrt{z \bar{z}}+\bar{z}-9\right)+2 z \left(\bar{z} \left(7-3 \sqrt{z \bar{z}}\right)+8 \sqrt{z \bar{z}}-7\right)-16 \sqrt{z \bar{z}}+9\right)+6 \sqrt{z \bar{z}}-1\right)\right.\nonumber\\
&&\left. -2 \sqrt{z \bar{z}}\right)).\nonumber\\ \label{ppoddblock}
\end{eqnarray}                           
Note that the singularity  structure for the parity odd case is  different from the parity even one due to the absence of $\log z $ terms in the expression. Here it is is due the presence of the square root branch cut. 
To see the origin of the singularity of going round the branch cut with the prescription $z\rightarrow z e^{-2i\pi}$,
let us examine the asymmetrically shifted scalar blocks in $(z,\bar{z})$.
\begin{eqnarray}
\lim_{\bar{z} \rightarrow 1}G^{(2,\Delta_\phi,\Delta_\phi,3)}(\tilde{u},\tilde{w}) &=&\frac{4 \left(\sqrt{z}-1\right)^2 \sqrt{(z-1) (\bar{z}-1)}}{\left(\sqrt{z}+1\right)^2 \sqrt{z}}.
\end{eqnarray} 
Under the transformation $z\rightarrow z e^{-2i\pi}$, the above expression picks up a pole on the second sheet, due to the square root in the denominator. We systematically apply this procedure to the parity odd blocks given in (\ref{ppoddblock}). 
 Parametrizing the coordinates $z, \bar{z}$ as given in equation  \eqref{coordparametrize}, we obtain the light cone limit of the parity odd sector 
\begin{eqnarray}\label{parityoddg}
\hat{G}^{x+}_\eta(\sigma)&=&\lim_{a \rightarrow \infty} \frac{8\sqrt{\eta } \lambda_{\phi\phi T} p_j}{ a^4 \sqrt{C_T} \sigma }\nonumber\\
\hat{G}^{+x}_\eta(\sigma)&=&\lim_{a \rightarrow \infty} \frac{8\sqrt{\eta } \lambda_{\phi\phi T} p_j}{ a^4 \sqrt{C_T} \sigma }\nonumber\\
\end{eqnarray}
This pole structure is independent of the way the differential operators are acted on the scalar block. We can perform the analytic continuation and then act on by the differential operators or vice-versa. We obtain the same result for both the cases. 
\subsection{Collider bounds}\label{colliderbounds}
                            \paragraph{}Now we have all the ingredients to formulate our bounds using the sum rule. Following \citep{Hartman:2015lfa, Rejon-Barrera:2015bpa}  we use the arguments of reflection positivity to bound the four point functions that we have obtained in the previous section. Let us begin by reviewing the sum rule that one gets for the scalar case \citep{Hartman:2015lfa}.
                            
\begin{eqnarray}\label{scalarcorr}
G^{s}_\eta(\sigma)&=& \frac{\langle \mathcal{O}(0)\phi(z,\bar{z}) \phi(1,1) \mathcal{O}(\infty))  \rangle}{\langle \phi(z,\bar{z}) \phi(1,1) \rangle},\nonumber\\
\hat{G}^{s}_\eta(\sigma)&=& \frac{\langle \mathcal{O}(0)\phi(ze^{-2\pi i},\bar{z}) \phi(1,1) \mathcal{O}(\infty))  \rangle}{\langle \phi(z,\bar{z}) \phi(1,1) \rangle}.\nonumber\\
\end{eqnarray}
              One can construct the following state in radial quantization by smearing the $\phi$ insertion over an unit disc \citep{Hartman:2015lfa}.
\begin{eqnarray}
|f\rangle &=& \int_0^1 dr_1 \int_0^{2\pi} d\theta_1 f(r,\theta_1) \phi(r_1 e^{i\theta_1},r_1 e^{-i\theta_1}) \mathcal{O}(0)|0\rangle.             
\end{eqnarray} 
Reflection positivity states that $\langle f|f \rangle \geq 0$. Reflection positivity of the correlator, eqn \eqref{scalarcorr}, in the s and u channels allows us to write the following sum rule for the t-channel when the exchanged object is a conserved stress tensor, 
\begin{eqnarray}
\text{Re}\left( G^s_{\eta}(\sigma) - \hat{G}^{s}_\eta(\sigma)\right) &\geq & 0,\nonumber\\
\oint_{\partial D} d\sigma \left( -G^s_{\eta}(\sigma) + \hat{G}^{s}_\eta(\sigma)\right) &=& 0,
\end{eqnarray}             
where the contour $D$ is the closed region spanned by a semicircle $S$ in the complex $\sigma$ plane of radius $R$ centering the origin.
 We can decompose the sum rule as follows 
\begin{eqnarray} \label{sumrule}
\text{Re}\int_{S} d\sigma\left( -G^s_{\eta}(\sigma) + \hat{G}^{s}_\eta(\sigma)\right) &=& \text{Re}\int_{-R}^{R} d\sigma\left( G^s_{\eta}(\sigma) - \hat{G}^{s}_\eta(\sigma)\right) \geq  0,
\end{eqnarray}  
where we have used the reflection positivity constraint in the second line. Note that the contour is to be traversed in $S$ is counter clockwise. The integral over $S$ can be now used to isolate the singularities in the $\hat{G}^s_{\eta}(\sigma)$. This works out in the following way. While leading terms of $G^s_{\eta}(\sigma)$ and $\hat{G}^s_{\eta}(\sigma)$ cancel each other, the subleading terms of $G^s_{\eta}(\sigma)$ are analytic in $\sigma$ but the first subleading term of $\hat{G}^s_{\eta}(\sigma)$ has a pole in $\sigma$. The integral precisely picks out residue of this pole and this leads to certain positivity constraints on the residue. Similar arguments holds for for the spinning external correlators. For reflection positivity, we consider the following states
\begin{eqnarray}\label{statenorm}
|f\rangle &=& \int_0^1 dr_1 \int_0^{2\pi} d\theta_1 f(r,\theta_1) \phi(r_1 e^{i\theta_1},r_1 e^{-i\theta_1}) j^\mu \epsilon_\mu (0)|0\rangle .            
\end{eqnarray}
The corresponding ket state involves the inversion operator $I^{\mu\nu}(x)= \eta^{\mu\nu} - 2 \frac{x^\mu x^\nu}{x^2}$ and this leads to some additional signs which one has to be careful about \citep{Li:2015itl}. Let us denote the state created by the $\epsilon_+ =1$ as $|+\rangle$ state and that created by $\epsilon_x =1$ as the $|x\rangle$ state. Reflection positivity implies that 
\begin{eqnarray}
\langle + | +\rangle \geq 0, \qquad \langle x|x \rangle \geq 0.
\end{eqnarray} 
This leads to the sum rules corresponding to eqn  \eqref{parityeveng} \citep{Li:2015itl}. We note that the reflection positivity stated above cannot capture the parity odd coefficient. This is due to the fact that the parity odd blocks do not contribute in the light cone limit for such a choice of polarizations. We now consider the positivity and analyticity of the following state $|v\rangle = \left(|+\rangle + \alpha |x\rangle\right)$. Reflection positivity implies,

\begin{eqnarray}
&&\langle v |v  \rangle\sim \langle +|+ \rangle + \alpha \langle +|x \rangle +\alpha^* \langle x| +\rangle + \alpha\alpha^* \langle x|x \rangle \geq 0,\nonumber\\
&&\sim \text{Re} \left( G^{vv}_{\eta}(\sigma) - \hat{G}^{vv}_\eta(\sigma)\right) \geq 0, \qquad \sigma\in[-R,R] .\nonumber\\
\end{eqnarray}
Reflection positivity also implies analyticity of $G^{++}$ and $G^{xx}$ correlator in the region D, 
\begin{eqnarray}
\oint_{\partial D}  d\sigma\left( G^{vv}_{\eta}(\sigma) - \hat{G}^{vv}_\eta(\sigma)\right)=0 .
\end{eqnarray}
Separating this integral over the semicircle and the real line, and considering the real part we obtain,
\begin{eqnarray}\label{causalitysumrule}
&&\text{Re}\left(-\int_s d\sigma\left( -G^{++}_{\eta}(\sigma) + \hat{G}^{++}_\eta(\sigma) \right)-\alpha\int_s d\sigma\left( -G^{+x}_{\eta}(\sigma) + \hat{G}^{+x}_\eta(\sigma) \right)\right.\nonumber\\
&&\left.  + \alpha^*\int_s d\sigma\left( -G^{x+}_{\eta}(\sigma) + \hat{G}^{x+}_\eta(\sigma) \right)+\alpha\alpha^*\int_s d\sigma\left( -G^{xx}_{\eta}(\sigma) + \hat{G}^{xx}_\eta(\sigma) \right)\right)\nonumber\\
&& = \text{Re}\left(-\int_{-R}^R d\sigma\left( G^{++}_{\eta}(\sigma) - \hat{G}^{++}_\eta(\sigma) \right) -\alpha\int_{-R}^{R} d\sigma\left( G^{+x}_{\eta}(\sigma) - \hat{G}^{+x}_\eta(\sigma) \right) \right.\nonumber\\
&&\left. + \alpha^*\int_{-R}^R d\sigma\left( G^{x+}_{\eta}(\sigma) - \hat{G}^{x+}_\eta(\sigma) \right) +\alpha\alpha^*\int_{-R}^R d\sigma\left( G^{xx}_{\eta}(\sigma) - \hat{G}^{xx}_\eta(\sigma) \right)\right)\geq 0 ,\nonumber\\
\end{eqnarray}
where the inequality in the final line is a consequence of reflection positivity for the $G^{vv}$ state.
Minimising with respect to $\alpha$, we obtain Cauchy-Schwartz inequality
\begin{eqnarray}\label{csineq}
\text{Re}\left(R^{++} R^{xx} -R^{+x}R^{x+} \right)\geq 0,
\end{eqnarray}
where $R^{ij}$s are described below.
\subsubsection*{No Subtraction}  
\begin{figure}[h]
\center
\includegraphics[scale=0.7]{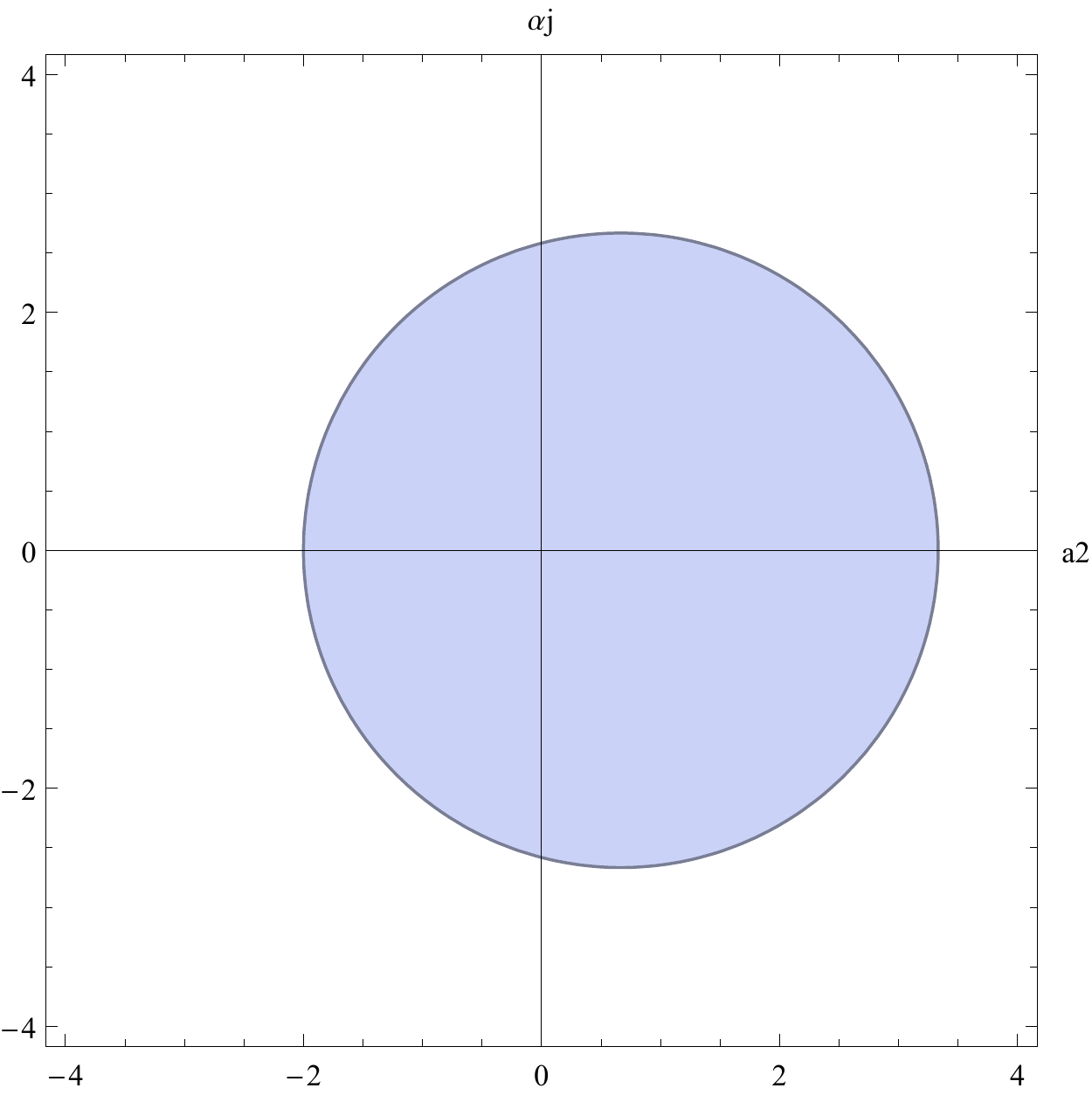}
\caption{Circle with no subtraction}\label{bound1}
\end{figure}

                 \paragraph{ } 
                 Let us proceed at first naively and 
                 and  assume no subtractions and not use the inner product constructed in (\ref{prescription}). 
                  From equation \eqref{parityevennosubg} and \eqref{parityoddg} we obtain 
\begin{eqnarray}
R^{++} &=& -\int_{S} d\sigma\left( -G^{++}_{\eta}(\sigma) + \hat{G}^{++}_\eta(\sigma)\right),\nonumber\\
&=& \frac{-32C_J \sqrt{\eta} \lambda_{\phi\phi T}(a_2+2)}{\pi \sqrt{C_T}},\nonumber\\
R^{+x} &=& -\int_{S} d\sigma\left( -G^{+x}_{\eta}(\sigma) + \hat{G}^{+x}_\eta(\sigma)\right),\nonumber\\
&=& \left( \frac{-\pi}{i} \right)\frac{\left( -8p_j \right) \lambda_{\phi\phi T} \sqrt{\eta}}{\sqrt{C_T}},\nonumber\\ 
R^{x+} &=& \int_{S} d\sigma\left( -G^{x+}_{\eta}(\sigma) + \hat{G}^{x+}_\eta(\sigma)\right),\nonumber\\
&=& \left( \frac{-\pi}{i} \right)\frac{\left( 8p_j \right) \lambda_{\phi\phi T} \sqrt{\eta}}{\sqrt{C_T}},\nonumber\\ 
R^{xx} &=& \int_{S} d\sigma\left( -G^{xx}_{\eta}(\sigma) + \hat{G}^{xx}_\eta(\sigma)\right),\nonumber\\
&=& \frac{32C_J \sqrt{\eta} \lambda_{\phi\phi T}(-10 + 3a_2)}{3\pi \sqrt{C_T}}.\nonumber\\
\end{eqnarray}
Note that $\lambda_{\phi\phi T}$ is negative and we have also scaled out the term $a^4$. 
Putting these back into eqn \eqref{csineq}, we obtain 
\begin{eqnarray}
\left(a_2 -\frac{2}{3}\right)^2 + \alpha_j^2 \leq \frac{64}{9},
\end{eqnarray}
where we have used
\begin{eqnarray}
\alpha_j^2 &=& \frac{\pi^4 p_j^2 }{16 C_J^2}.
\end{eqnarray}
This bound is  drawn in figure  \eqref{bound1}
The bounds that we get  proceeding naively  are not optimal and do not coincide with that obtained
using the average null energy condition. 
We do not consider the subtractions 
as outlined in section \ref{jjphiphifromphi}.
\newpage
      

\subsubsection*{Optimal bound}
\begin{figure}[h]
\center
\includegraphics[scale=0.7]{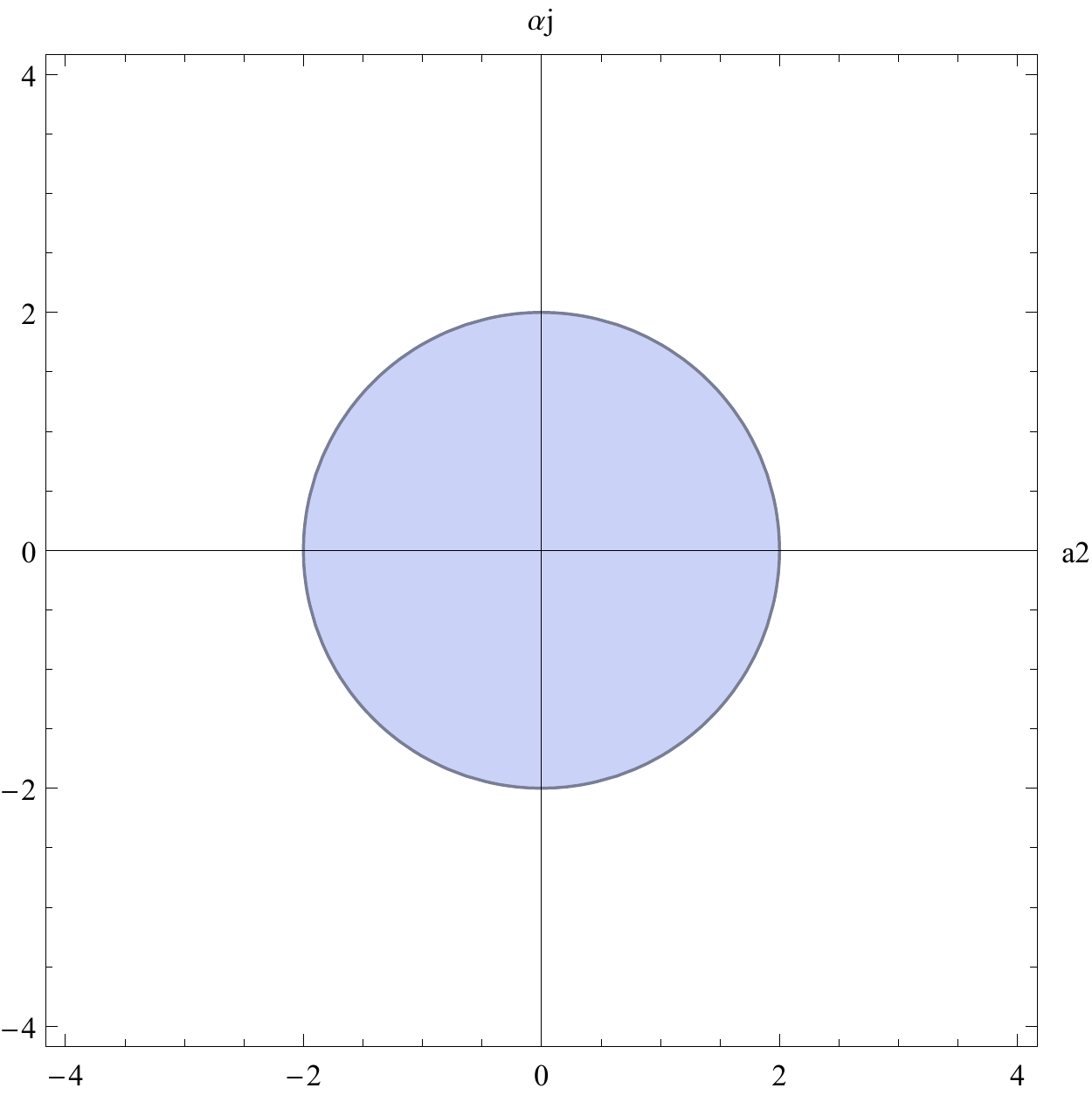}
\caption{Optimal bounds}\label{bound3}
\end{figure}

                      The optimal bound is obtained by taking the subtracted 
                     parity even $\tilde{G}^{xx}$ given in eqn \eqref{parityeveng}. Our inequality therefore becomes
\begin{eqnarray}
\text{Re}\left(R^{++} \tilde{R}^{xx} -R^{+x}R^{x+} \right)\geq 0,
\end{eqnarray} 
where the individual components are described as, 
              \begin{eqnarray}
R^{++} &=& -\int_{S} d\sigma\left( -G^{++}_{\eta}(\sigma) + \hat{G}^{++}_\eta(\sigma)\right),\nonumber\\
&=& \frac{-32C_J \sqrt{\eta} \lambda_{\phi\phi T}(a_2+2)}{\pi \sqrt{C_T}},\nonumber\\
R^{+x} &=& -\int_{S} d\sigma\left( -G^{+x}_{\eta}(\sigma) + \hat{G}^{+x}_\eta(\sigma)\right),\nonumber\\
&=& \left( \frac{-\pi}{i} \right)\frac{\left( -8p_j \right) \lambda_{\phi\phi T} \sqrt{\eta}}{\sqrt{C_T}},\nonumber\\ 
R^{x+} &=& \int_{S} d\sigma\left( -G^{+x}_{\eta}(\sigma) + \hat{G}^{+x}_\eta(\sigma)\right),\nonumber\\
&=& \left( \frac{-\pi}{i} \right)\frac{\left( 8p_j \right) \lambda_{\phi\phi T} \sqrt{\eta}}{\sqrt{C_T}},\nonumber\\ 
\tilde{R}^{xx} &=& \int_{S} d\sigma\left( -\tilde{G}^{xx}_{\eta}(\sigma) + \hat{\tilde{G}}^{xx}_\eta(\sigma)\right),\nonumber\\
&=& \frac{32C_J \sqrt{\eta} \lambda_{\phi\phi T}(a_2-2)}{\pi \sqrt{C_T}}.\nonumber\\
\end{eqnarray}
Note that $\lambda_{\phi\phi T}$ is negative and as one would expect for a theory with no parity violating terms,  $\tilde{R}^{xx}$ is positive. Putting these back into eqn \eqref{csineq}, we obtain 
\begin{eqnarray}
a_2^2 + \alpha_j^2 \leq 4.
\end{eqnarray}
This can be pictorially represented in figure \ref{bound3}
 
These observations can be summarized in the following figure
\ref{bound4}.
\begin{figure}[h]
\center
\includegraphics[scale=0.7]{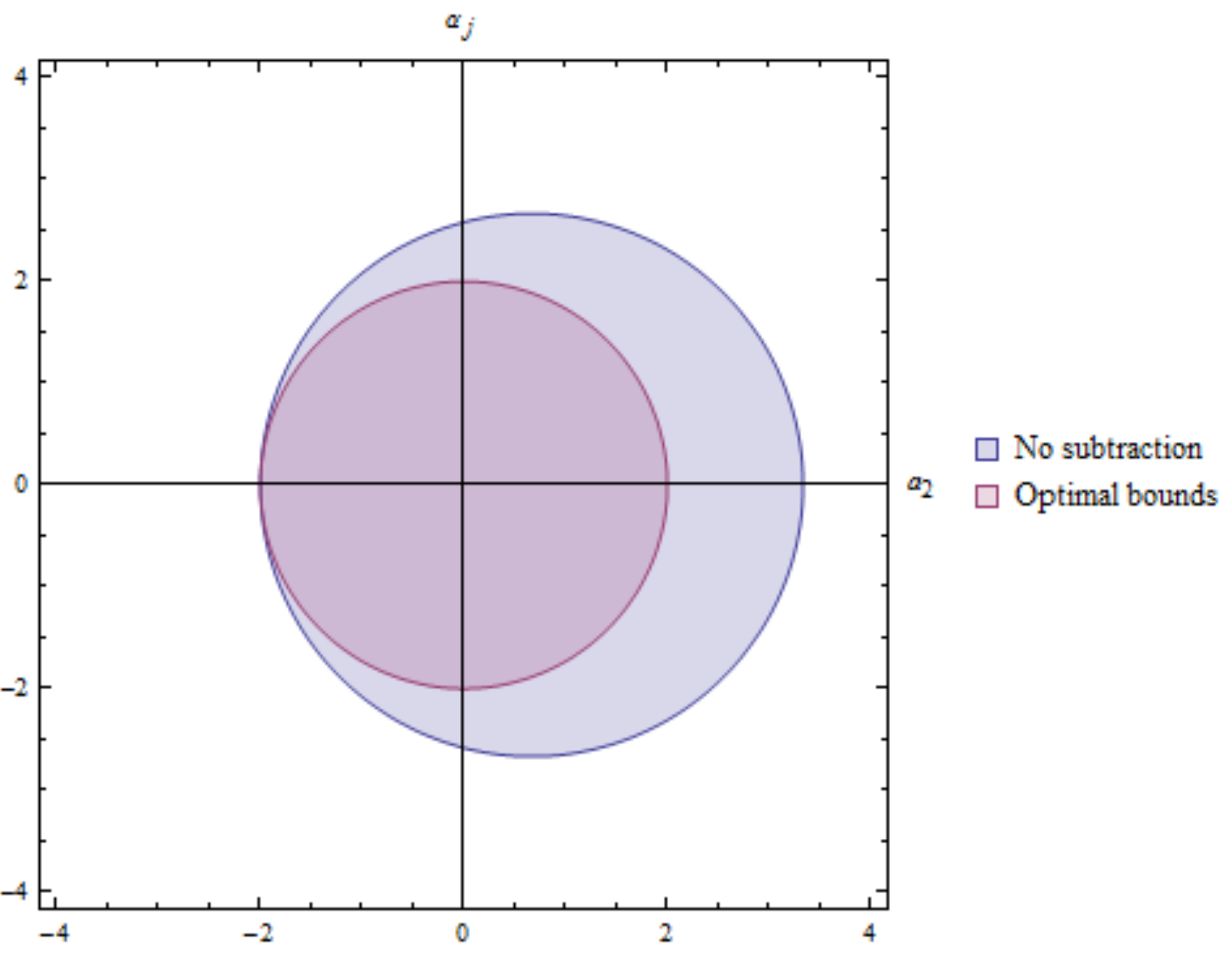}
\caption{Comparative study of bounds}\label{bound4}
\end{figure} 

Note that we have obtained 
 the causality sum rules eqn  \eqref{causalitysumrule} for different subtraction schemes in this section. We find that the  optimal bounds coincide with the collider constraints evaluated in \cite{Chowdhury:2017vel}. 
 Obtaining this involved   the subtraction symmetric traceless tensor  exchange in 
 the $(xx)$ component of the parity even contributions as well as a normalisation as given in 
 (\ref{prescription}).  Reflection positivity holds even after subtraction since all we have done by the subtraction 
 is to isolated the anti-symmetric tensor exchange. 
We would like to have a more first principle method of arriving at the normalisation used in 
(\ref{prescription}).   Performing the same analysis for the $\langle TT\phi\phi \rangle$ to obtain similar 
bounds on the OPE coefficients of the stress tensor will give us a clue. 

Finally we note  the fundamental difference between the nature of the singularity of the light cone block on the second sheet for the parity even case and the parity odd case. While for the parity even conformal block, it results from the logarithm of the hypergeometric function, for the parity odd case this results from a square root branch cut. The absence of logarithms at this order in the light cone expansion 
 in the parity odd conformal block can be technically understood as due to the asymmetric shifts in the scalar blocks due to the $D_{\text{odd}}$ operators of equation \eqref{dodd}. This is also manifests itself in modifications to the crossing equation, the parity odd corrections to the stress tensor exchange in the $s$-channel give rise to new OPE coefficients \eqref{mixedope} corresponding to a new class of double twist operators rather than anomalous dimensions.

\section{Conclusions}
\paragraph{ } In this paper we have studied the modifications to the crossing equation for a four point function of two $U(1)$ currents and two scalars due to the presence of a parity violating term in the stress tensor exchange in the $s$-channel. We find that for the crossing equations to be satisfied, there exists a tower of new double trace operators in the $t$-channel. We find the requisite differential operators which can give rise to such operator exchanges. An infinite sum over such double trace operators reproduces the parity odd contribution in the $s$-channel. We find that due to the structure of the parity odd blocks, such an exchange contributes to the OPE coefficients  rather than the anomalous dimensions of operators in the t-channel. We also study constraints imposed by causality considerations on such a four point function. We find that using reflection positivity and crossing symmetry we can formulate a Cauchy Schwartz inequality which leads to the exact parity odd conformal collider bounds that was studied previously \cite{Chowdhury:2017vel}. For this purpose it was necessary to look at the parity odd blocks in the light cone limit after certain analytic continuations. The nature of singularity on the second sheet for the parity odd blocks is fundamentally different from that of the parity even blocks \citep{Hartman:2015lfa, Hofman:2016awc} even though the singularity structure is the same. This again boils down to the differences in the structure of the party odd and parity even conformal blocks which was responsible for absence of anomalous dimensions in the crossed channel. Our results  provide insight on 
how 
the constraints imposed on parity odd three point function parameters by conformal collider physics 
emerge from reflection positivity, analyticity and crossing symmetry. The fact that the bounds seen in collider 
physics can be obtained this way 
 resonates with the fact that any causal theory including parity odd ones must satisfy 
 average null energy condition \cite{Hartman:2016lgu} and hence the collider bounds.
\paragraph{ } There are several future directions. A naive generalization of this analysis to a correlator of two stress tensors and two scalars should lead to conformal collider constraints on the parity violating three point function of the stress tensors. Moreover similar to the four point function considered in this paper, the crossing equation of the four point function of stress tensors and scalars will be modified due to the presence of the parity violating term in the stress tensor exchange in the $s$-channel. Differential operators for stress tensor exchange in four point functions of four $U(1)$ currents also have this parity odd term which has not been considered before \cite{Li:2015itl}. It will be interesting to see the modifications to the crossing equation of this correlator because of the presence of the parity odd term.        
 
\section{Acknowledgements}
The authors would like to thank Tom Hartman and David Simmons-Duffin for comments on the draft. 
S.D.C would like to thank the Simons Bootstrap collaboration for organising the Bootstrap School 2018, for providing hospitality during the project and Kausik Ghosh for discussions. The research
was supported in part by the International Centre for
Theoretical Sciences (ICTS) during a visit for participating in the
program - AdS/CFT at 20 and Beyond (Code: ICTS/adscft20/05). The work
of S.P. was supported in part by a DST-SERB Early Career Research
Award (ECR/2017/001023) and a DST INSPIRE Faculty
Award.
\appendix

\section{Differential operators on RHS for stress tensor exchange}\label{lhsdiffops}
                                                    
                        \paragraph{ } The spinning conformal blocks have been expressed in terms of differential operators acting on the scalar blocks eqn \eqref{scblock}. There are two contributions to the final result. Explicit action of the differential blocks on the scalar blocks can be grouped into parity even and parity odd contributions.
\subsection{Parity even}                          
                         \begin{eqnarray}\label{parityevendifflhs}
D_{\text{even}}  {\cal W}(2, 2, \Delta_\phi, \Delta_\phi)&=& \alpha_3 \left(\left(2 \lambda_{jjT} -\frac{3 C_J}{8 \pi }\right)D_{11}D_{22} +  \left(2 \lambda_{jjT} -\frac{9 C_J}{8 \pi }\right)D_{12}D_{21}-2 \lambda_{jjT} H_{12}\right)\nonumber\\
&&\times \sum^{1,1} {\cal W}(2, 2, \Delta_\phi, \Delta_\phi)\nonumber\\
&=&  \left(\left(2 \lambda_{jjT} -\frac{3 C_J}{8 \pi }\right)I^{\text{even}}_1 +  \left(2 \lambda_{jjT} -\frac{9 C_J}{8 \pi }\right)I^{\text{even}}_2 -2 \lambda_{jjT} I^{\text{even}}_3\right)\nonumber\\
\end{eqnarray}
where the terms $I^{\text{even}}_i$s are given by
\begin{eqnarray}
I^{\text{even}}_1 &=& D_{11}D_{22}\sum^{1,1} {\cal W}(2, 2, \Delta_\phi, \Delta_\phi)\nonumber\\
&=&\frac{\alpha_3}{(-2)^{3+\Delta_\phi}} \frac{1}{(P_1.P_2)^3(P_3.P_4)^{\Delta_\phi}} \left( \left( V_{1,23} V_{2,13} + V_{1,24}V_{2,14}\right)uw\partial_u\partial_w + V_{1,23}V_{2,14}(w\partial_w)^2 \right.\nonumber\\
&&\left.+ V_{1,24}V_{2,13}(u\partial_u)^2 -\frac{1}{2}H_{12} \left(u\partial_u + w\partial_w \right)\right)G^{3,3,\Delta_\phi,\Delta_\phi}_{{\cal O}}(u,w)
\end{eqnarray}
\begin{eqnarray}
I^{\text{even}}_2 &=& D_{12}D_{21}\sum^{1,1} {\cal W}(2, 2, \Delta_\phi, \Delta_\phi)\nonumber\\
&=&\frac{\alpha_3}{(-2)^{3+\Delta_\phi}} \frac{1}{(P_1.P_2)^3(P_3.P_4)^{\Delta_\phi}} \left( \left( V_{1,23} V_{2,13} + V_{1,24}V_{2,14}\right)uw\partial_u\partial_w + \frac{wV_{2,13}V_{1,24}}{u}(w\partial_w)^2 \right.\nonumber\\
&&\left.+ \frac{uV_{2,14}V_{1,23}}{w}(u\partial_u)^2 -\frac{1}{2}H_{12} \left(u\partial_u + w\partial_w \right)\right)G^{3,3,\Delta_\phi,\Delta_\phi}_{{\cal O}}(u,w)
\end{eqnarray}

\begin{eqnarray}
I^{\text{even}}_3 &=& H_{12}\sum^{1,1} {\cal W}(2, 2, \Delta_\phi, \Delta_\phi)\nonumber\\
&=& \frac{\alpha_3}{(-2)^{3+\Delta_\phi}}\frac{H_{12}}{(P_1.P_2)^3(P_3.P_4)^{\Delta_\phi}}G^{3,3,\Delta_\phi,\Delta_\phi}_{{\cal O}}(u,w)\nonumber\\
\end{eqnarray}

\subsection{Parity odd}\label{parityodddifflhs2}
\begin{eqnarray}\label{parityodddifflhs}
D_{\text{odd}} {\cal W}(2,2, \Delta_\phi, \Delta_\phi) &=&\frac{-i p_j}{1152}  \left({\cal D}^{(3)} + {\cal D}^{(4)}\right){\cal W}(2,2, \Delta_\phi, \Delta_\phi)\nonumber\\
&=& \frac{-i p_j}{1152}  \left( I^{\text{odd}}_1 + I^{\text{odd}}_2 + I^{\text{odd}}_3 + I^{\text{odd}}_4 \right)\nonumber\\
\end{eqnarray}
Where the terms $I^{\text{odd}}_i$ are given by,
\begin{eqnarray}
I^{\text{odd}}_1 &=&  8\tilde{D}_1D_{21}\sum^{1,0}{\cal W}(2, 2, \Delta_\phi, \Delta_\phi)\nonumber\\
&=&(-\frac{2\epsilon_{abcde}Z_1^a P_1^b P_3^c P_2^d P_4^e}{\left(P_1.P_2\right){}^{5/2} P_1.P_3 \left(P_2.P_3\right){}^2 \left(P_2.P_4\right){}^3}\left(\frac{P_2.P_4}{P_1.P_4}\right){}^{3/2}\left(P_3.P_4\right){}^{-\Delta_\phi}  \nonumber\\
&&\left(w P_1.P_3 P_2.P_4 \left(P_2.P_3 P_2.P_4 P_1.Z_2 \left(3 \partial_w  G^{(3,2,\Delta_\phi,\Delta_\phi)}(u,w)+6 u \partial_u \partial_w  G^{(3,2,\Delta_\phi,\Delta_\phi)}(u,w)\right.\right.\right.\nonumber\\
&&\left. \left. \left. +4 w \left(3 \partial_w^2  G^{(3,2,\Delta_\phi,\Delta_\phi)}(u,w)\right.\right.\right.\right.\nonumber\\
&&\left. \left. \left. \left. +w \partial_w^3  G^{(3,2,\Delta_\phi,\Delta_\phi)}(u,w)+u \partial_u \partial_w^2  G^{(3,2,\Delta_\phi,\Delta_\phi)}(u,w)\right)\right)-P_1.P_2 \left(2 P_2.P_4 P_3.Z_2\right.\right.\right.\nonumber\\
&&\left.\left.\left. \left(2 w^2 \partial_w^3  G^{(3,2,\Delta_\phi,\Delta_\phi)}(u,w)+9 w \partial_w^2  G^{(3,2,\Delta_\phi,\Delta_\phi)}(u,w)\right.\right.\right.\right.\nonumber\\
&&\left.\left.\left.\left. +6 \partial_w  G^{(3,2,\Delta_\phi,\Delta_\phi)}(u,w)\right)+P_2.P_3 P_4.Z_2 \left(-3 \partial_w  G^{(3,2,\Delta_\phi,\Delta_\phi)}(u,w)-2 w \partial_w^2  G^{(3,2,\Delta_\phi,\Delta_\phi)}(u,w)\right.\right.\right.\right.\nonumber\\
&&\left.\left.\left.\left. +6 u \partial_u \partial_w  G^{(3,2,\Delta_\phi,\Delta_\phi)}(u,w)\right.\right.\right.\right.\nonumber\\
&&\left.\left.\left.\left. +4 u w \partial_u \partial_w^2  G^{(3,2,\Delta_\phi,\Delta_\phi)}(u,w)\right)\right)\right)+uP_1.P_4 P_2.P_3  \left(P_1.P_2 \left(2 P_2.P_4 w P_3.Z_2 \left(\partial_u \partial_w  G^{(3,2,\Delta_\phi,\Delta_\phi)}(u,w)\right.\right.\right.\right.\nonumber\\
&&\left.\left.\left.\left.+2 u \partial_u^2 \partial_w  G^{(3,2,\Delta_\phi,\Delta_\phi)}(u,w)\right) +P_2.P_3 P_4.Z_2 \left(3 \partial_u G^{(3,2,\Delta_\phi,\Delta_\phi)}(u,w)+4 u \left(3 \partial_u^2G^{(3,2,\Delta_\phi,\Delta_\phi)}(u,w)\right.\right.\right.\right.\right.\nonumber\\
&& \left.\left.\left.\left.\left. +u \partial_u^3G^{(3,2,\Delta_\phi,\Delta_\phi)}(u,w)\right)\right)\right) \right.\right.\nonumber\\
&&\left.\left. -P_2.P_3 P_2.P_4 P_1.Z_2 \left(\partial_u G^{(3,2,\Delta_\phi,\Delta_\phi)}(u,w)+2 w \partial_u \partial_w  G^{(3,2,\Delta_\phi,\Delta_\phi)}(u,w)\right.\right.\right.\nonumber\\
&&\left.\left.\left. +4 u \left(2 \partial_u^2G^{(3,2,\Delta_\phi,\Delta_\phi)}(u,w)\right.\right.\right.\right.\nonumber\\
&& \left.\left.\left.\left. +w \partial_u^2 \partial_w  G^{(3,2,\Delta_\phi,\Delta_\phi)}(u,w) +u \partial_u^3G^{(3,2,\Delta_\phi,\Delta_\phi)}(u,w)\right)\right)\right)\right)\nonumber\\
&&+\left(-\frac{8 \epsilon_{abcde}Z_2^a P_1^b P_3^c P_2^d Z_1^e }{\left(P_1.P_2\right){}^{5/2} P_2.P_3}\sqrt{\frac{P_2.P_4}{P_1.P_4}}\left(P_3.P_4\right){}^{-\Delta_\phi}w\partial_w  G^{(3,2,\Delta_\phi,\Delta_\phi)}(u,w)\right)\nonumber\\
&&+\left( \frac{4 \epsilon_{abcde} Z_2^a P_1^b P_4^c P_2^d Z_1^e }{\left(P_1.P_2\right){}^{5/2} P_2.P_4}\sqrt{\frac{P_2.P_4}{P_1.P_4}}\left(P_3.P_4\right){}^{-\Delta_\phi} \left(G^{(3,2,\Delta_\phi,\Delta_\phi)}(u,w)-2 u \partial_u  G^{(3,2,\Delta_\phi,\Delta_\phi)}(u,w)\right)  \right)\nonumber\\
&&+\left( \frac{4 \epsilon_{abcde} Z_2^a P_1^b P_4^c P_3^d Z_1^e   }{\left(P_1.P_2\right){}^{3/2} P_1.P_3 \left(P_1.P_4\right){}^2 P_2.P_3 \sqrt{\frac{P_2.P_4}{P_1.P_4}}}\right.\nonumber\\
&&\left.\left(P_3.P_4\right){}^{-\Delta_\phi}\left(P_1.P_3 P_2.P_4 w \left(3 \partial_w G^{(3,2,\Delta_\phi,\Delta_\phi)}(u,w)+2 w \partial_w^2 G^{(3,2,\Delta_\phi,\Delta_\phi)}(u,w)\right)\right.\right.\nonumber\\
&&\left.\left. -P_1.P_4 P_2.P_3 u \left(\partial_u G^{(3,2,\Delta_\phi,\Delta_\phi)}(u,w)  +2 u \partial_u^2 G^{(3,2,\Delta_\phi,\Delta_\phi)}(u,w)\right)\right) \right)\nonumber\\
&&\times \frac{1}{(-2)^{\frac{5}{2}+\Delta_\phi}}
\nonumber\\
\nonumber\\
\end{eqnarray}
\begin{eqnarray}
I^{\text{odd}}_2 &=&  4\tilde{D}_1D_{22}\sum^{0,1}{\cal W}(2, 2, \Delta_\phi, \Delta_\phi)\nonumber\\
&=&(\frac{\epsilon_{abcde}Z_1^a P_1^b P_3^c P_2^d P_4^e  }{\left(P_1.P_2\right){}^{5/2} \left(P_1.P_3\right){}^2 \left(P_1.P_4\right){}^3 P_2.P_3}\left(\frac{P_1.P_4}{P_2.P_4}\right){}^{3/2}\left(P_3.P_4\right){}^{-\Delta_\phi} \left(P_2.P_4 \left(P_1.P_3\right){}^2 (-w) \right. \nonumber\\
&&\left. \left(P_2.P_4 P_1.Z_2-P_1.P_2 P_4.Z_2\right) \left(3 \partial_w  G^{(2,3,\Delta_\phi, \Delta_\phi)}(u,w)+4 w \left(3 \partial_w^2  G^{(2,3,\Delta_\phi, \Delta_\phi)}(u,w)\right.\right.\right.\nonumber\\
&&\left.\left.\left. +w \partial_w^3  G^{(2,3,\Delta_\phi, \Delta_\phi)}(u,w)\right)\right)\right.\nonumber\\
&& \left. +P_1.P_4 P_1.P_3 u \left(P_2.P_3 P_2.P_4 P_1.Z_2 \left(-3 \partial_u  G^{(2,3,\Delta_\phi, \Delta_\phi)}(u,w)-2 u \partial_u^2 G^{(2,3,\Delta_\phi, \Delta_\phi)}(u,w)\right.\right.\right.\nonumber\\
&&\left.\left.\left.+4 w \left(\partial_u \partial_w G^{(2,3,\Delta_\phi, \Delta_\phi)}(u,w)\right.\right.\right.\right.\nonumber\\
&&\left.\left.\left.\left.-w \partial_u \partial_w^2 G^{(2,3,\Delta_\phi, \Delta_\phi)}(u,w)+u \partial_u^2 \partial_w G^{(2,3,\Delta_\phi, \Delta_\phi)}(u,w)\right)\right)+P_1.P_2 \left(2 P_2.P_4 w P_3.Z_2 \right.\right.\right.\nonumber\\
&&\left.\left.\left.\left(\partial_u \partial_w G^{(2,3,\Delta_\phi, \Delta_\phi)}(u,w)\right.\right.\right.\right.\nonumber\\
&&\left.\left.\left.\left. +2 w \partial_u \partial_w^2 G^{(2,3,\Delta_\phi, \Delta_\phi)}(u,w)\right)+P_2.P_3 P_4.Z_2 \left(3 \partial_u  G^{(2,3,\Delta_\phi, \Delta_\phi)}(u,w)\right.\right.\right.\right.\nonumber\\
&&\left.\left.\left.\left.-6 w \partial_u \partial_w G^{(2,3,\Delta_\phi, \Delta_\phi)}(u,w)+2 u \partial_u^2 G^{(2,3,\Delta_\phi, \Delta_\phi)}(u,w) \right.\right.\right.\right.\nonumber\\
&&\left.\left.\left.\left.-4 u w \partial_u^2 \partial_w G^{(2,3,\Delta_\phi, \Delta_\phi)}(u,w)\right)\right)\right)+2 \left(P_1.P_4\right){}^2 P_2.P_3 u \left(P_2.P_3 P_1.Z_2-P_1.P_2 P_3.Z_2\right)\right.\nonumber\\
&&\left. \left(6 \partial_u  G^{(2,3,\Delta_\phi, \Delta_\phi)}(u,w) \right.\right.\nonumber\\
&&\left. \left. +u \left(9 \partial_u^2 G^{(2,3,\Delta_\phi, \Delta_\phi)}(u,w)+2 u \partial_u^3  G^{(2,3,\Delta_\phi, \Delta_\phi)}(u,w)\right)\right)\right))\nonumber\\
&&+\left(\frac{2 \epsilon_{abcde} Z_2^a P_1^b P_3^c P_2^d Z_1^e   }{\left(P_1.P_2\right){}^{5/2} P_1.P_3 P_1.P_4 P_2.P_3}\sqrt{\frac{P_1.P_4}{P_2.P_4}}\left(P_3.P_4\right){}^{-\Delta_\phi}\right.\nonumber\\
&&\left. \left(P_1.P_4 P_2.P_3 u \left(\partial_u  G^{(2,3,\Delta_\phi, \Delta_\phi)}(u,w)+2 u \partial_u^2  G^{(2,3,\Delta_\phi, \Delta_\phi)}(u,w)\right)-P_1.P_3 P_2.P_4 w \right.\right. \nonumber\\
&&\left.\left.\left(\partial_w  G^{(2,3,\Delta_\phi, \Delta_\phi)}(u,w)+2 w \partial_w^2  G^{(2,3,\Delta_\phi, \Delta_\phi)}(u,w)\right)\right)\right)\nonumber\\
&&+ \left( \frac{2 \epsilon_{abcde} Z_2^a P_1^b P_4^c P_2^d Z_1^e }{\left(P_1.P_2\right){}^{5/2} P_1.P_3 \sqrt{\frac{P_1.P_4}{P_2.P_4}} \left(P_2.P_4\right){}^2}\left(P_3.P_4\right){}^{-\Delta_\phi} \left(P_1.P_3 P_2.P_4 \left(w \left(2 w \partial_w^2 G^{(2,3,\Delta_\phi, \Delta_\phi)}(u,w) \right.\right.\right.\right.\nonumber\\
&&\left.\left.\left.\left. -\partial_w G^{(2,3,\Delta_\phi, \Delta_\phi)}(u,w)\right)+G^{(2,3,\Delta_\phi, \Delta_\phi)}(u,w)\right)-P_1.P_4 P_2.P_3 u \left(3 \partial_u G^{(2,3,\Delta_\phi, \Delta_\phi)}(u,w)\right.\right.\right. \nonumber\\
&&\left.\left.\left. +2 u \partial_u^2 G^{(2,3,\Delta_\phi, \Delta_\phi)}(u,w)\right)\right) \right)\nonumber\\
&&\times \frac{1}{(-2)^{\frac{5}{2}+\Delta_\phi}}\nonumber\\
\end{eqnarray}
\begin{eqnarray}
I^{\text{odd}}_3 &=& 8\tilde{D}_2D_{12}\sum^{0,1}{\cal W}(2, 2, \Delta_\phi, \Delta_\phi)\nonumber\\ 
&=& \left( \frac{2 \epsilon_{abcde}Z_2^a P_1^b P_3^c P_2^d P_4^e  }{\left(P_1.P_2\right){}^{5/2} \left(P_1.P_3\right){}^2 \left(P_1.P_4\right){}^3 P_2.P_3}\left(\frac{P_1.P_4}{P_2.P_4}\right){}^{3/2}\left(P_3.P_4\right){}^{-\Delta_\phi} \right.\nonumber\\
&&\left. \left(P_2.P_4 \left(P_1.P_3\right){}^2 (-w) \left(P_1.P_4 P_2.Z_1 \left(\partial_w  G^{(2,3,\Delta_\phi,\Delta_\phi)}(u,w)+2 u \partial_u \partial_w  G^{(2,3,\Delta_\phi,\Delta_\phi)}(u,w)\right.\right.\right.\right.\nonumber\\
&&\left.\left.\left.\left.+4 w \left(2 \partial_w^2  G^{(2,3,\Delta_\phi,\Delta_\phi)}(u,w) \right.\right.\right.\right.\right.\nonumber\\
&&\left.\left.\left.\left.\left. +w \partial_w^3  G^{(2,3,\Delta_\phi,\Delta_\phi)}(u,w)+u \partial_u \partial_w^2  G^{(2,3,\Delta_\phi,\Delta_\phi)}(u,w)\right)\right)-P_1.P_2 P_4.Z_1 \left(3 \partial_w  G^{(2,3,\Delta_\phi,\Delta_\phi)}(u,w)\right.\right.\right.\right.\nonumber\\
&&\left.\left.\left.\left. +4 w \left(3 \partial_w^2  G^{(2,3,\Delta_\phi,\Delta_\phi)}(u,w)+w \partial_w^3  G^{(2,3,\Delta_\phi,\Delta_\phi)}(u,w)\right)\right)\right)+P_1.P_4 P_1.P_3 u \left(P_1.P_2\right.\right.\right.\nonumber\\
&&\left.\left.\left. \left(2 P_2.P_4 w P_3.Z_1 \left(\partial_u \partial_w  G^{(2,3,\Delta_\phi,\Delta_\phi)}(u,w)\right.\right.\right.\right.\right.\nonumber\\
&&\left.\left.\left.\left.\left.+2 w \partial_u \partial_w^2  G^{(2,3,\Delta_\phi,\Delta_\phi)}(u,w)\right)+P_2.P_3 P_4.Z_1 \left(3 \partial_u G^{(2,3,\Delta_\phi,\Delta_\phi)}(u,w)-6 w \partial_u \partial_w  G^{(2,3,\Delta_\phi,\Delta_\phi)}(u,w)\right.\right.\right.\right.\right.\nonumber\\
&&\left.\left.\left.\left.\left. +2 u \partial_u^2 G^{(2,3,\Delta_\phi,\Delta_\phi)}(u,w)\right.\right.\right.\right.\right.\nonumber\\
&&\left.\left.\left.\left.\left.-4 u w \partial_u^2 \partial_w  G^{(2,3,\Delta_\phi,\Delta_\phi)}(u,w)\right)\right)+P_1.P_4 P_2.P_3 P_2.Z_1 \left(3 \partial_u G^{(2,3,\Delta_\phi,\Delta_\phi)}(u,w)\right.\right.\right.\right.\nonumber\\
&&\left.\left.\left.\left. +6 w \partial_u \partial_w  G^{(2,3,\Delta_\phi,\Delta_\phi)}(u,w)\right.\right.\right.\right.\nonumber\\
&&\left.\left.\left.\left. +4 u \left(3 \partial_u^2 G^{(2,3,\Delta_\phi,\Delta_\phi)}(u,w)+w \partial_u^2 \partial_w  G^{(2,3,\Delta_\phi,\Delta_\phi)}(u,w)+u \partial_u^3 G^{(2,3,\Delta_\phi,\Delta_\phi)}(u,w)\right)\right)\right)\right.\right.\nonumber\\
&&\left.\left.-2 P_1.P_2 \left(P_1.P_4\right){}^2 P_2.P_3 u P_3.Z_1 \left(6 \partial_u G^{(2,3,\Delta_\phi,\Delta_\phi)}(u,w)+u \left(9 \partial_u^2 G^{(2,3,\Delta_\phi,\Delta_\phi)}(u,w)\right.\right.\right.\right.\nonumber\\
&&\left.\left.\left.\left. +2 u \partial_u^3 G^{(2,3,\Delta_\phi,\Delta_\phi)}(u,w)\right)\right)\right)
\right)\nonumber\\
&&+\left( -\frac{8 \epsilon_{abcde} Z_2^a P_1^b P_3^c P_2^d Z_1^e }{\left(P_1.P_2\right){}^{5/2} P_1.P_3}\sqrt{\frac{P_1.P_4}{P_2.P_4}}\left(P_3.P_4\right){}^{-\Delta_\phi} u\partial_u  G^{(2,3,\Delta_\phi,\Delta_\phi)}(u,w) \right)\nonumber\\
&&+\left(\frac{4 \epsilon_{abcde} Z_2^a P_1^b P_4^c P_2^d Z_1^e   }{\left(P_1.P_2\right){}^{5/2} P_1.P_4}\sqrt{\frac{P_1.P_4}{P_2.P_4}}\left(P_3.P_4\right){}^{-\Delta_\phi} \left(G^{(2,3,\Delta_\phi,\Delta_\phi)}(u,w)\right.\right.\nonumber\\
&&\left. \left. -2 w \partial_w  G^{(2,3,\Delta_\phi,\Delta_\phi)}(u,w)\right) \right)\nonumber\\
&&+ \left(\frac{4 \epsilon_{abcde} Z_2^a P_2^b P_4^c P_3^d Z_1^e }{\left(P_1.P_2\right){}^{3/2} P_1.P_3 P_2.P_3 \sqrt{\frac{P_1.P_4}{P_2.P_4}} \left(P_2.P_4\right){}^2}\left(P_3.P_4\right){}^{-\Delta_\phi} \left(P_1.P_3 P_2.P_4 w \left(\partial_w  G^{(2,3,\Delta_\phi,\Delta_\phi)}(u,w)\right.\right.\right.\nonumber\\
&&\left.\left.\left. +2 w \partial_w^2  G^{(2,3,\Delta_\phi,\Delta_\phi)}(u,w)\right)-P_1.P_4 P_2.P_3 u \left(3 \partial_u  G^{(2,3,\Delta_\phi,\Delta_\phi)}(u,w)+2 u \partial_u^2  G^{(2,3,\Delta_\phi,\Delta_\phi)}(u,w)\right)\right)\right)\nonumber\\
&&\times \frac{1}{(-2)^{\frac{5}{2}+\Delta_\phi}}\nonumber\\
\end{eqnarray}
\begin{eqnarray}
I^{\text{odd}}_4
&=&4\tilde{D}_2D_{11}\sum^{1,0}{\cal W}(2, 2, \Delta_\phi, \Delta_\phi)\nonumber\\ 
&=& \left( \frac{ \epsilon_{abcde} Z_2^a P_1^b P_3^c P_2^d P_4^e}{\left(P_1.P_2\right){}^{5/2} P_1.P_3 \left(P_2.P_3\right){}^2 \left(P_2.P_4\right){}^3}\left(\frac{P_2.P_4}{P_1.P_4}\right){}^{3/2} \left(P_3.P_4\right){}^{-\Delta_\phi} \right.\nonumber\\
&&\left. \left(P_1.P_3 P_2.P_4 w \left(P_1.P_2 \left(2 P_2.P_4 P_3.Z_1 \left(2 w^2 \partial_w^3 G^{(3,2,\Delta_\phi,\Delta_\phi)}(u,w)+9 w \partial_w^2 G^{(3,2,\Delta_\phi,\Delta_\phi)}(u,w)\right.\right.\right.\right.\right.\nonumber\\
&&\left.\left.\left.\left.\left. +6 \partial_w G^{(3,2,\Delta_\phi,\Delta_\phi)}(u,w)\right)\right.\right.\right.\right.\nonumber\\
&&\left.\left.\left.\left. +P_2.P_3 P_4.Z_1 \left(-3 \partial_w G^{(3,2,\Delta_\phi,\Delta_\phi)}(u,w)-2 w \partial_w^2 G^{(3,2,\Delta_\phi,\Delta_\phi)}(u,w)+6 u \partial_u \partial_w G^{(3,2,\Delta_\phi,\Delta_\phi)}(u,w) \right.\right.\right.\right.\right.\nonumber\\
&&\left.\left.\left.\left.\left. +4 u w \partial_u \partial_w^2 G^{(3,2,\Delta_\phi,\Delta_\phi)}(u,w)\right)\right) \right.\right.\right.\nonumber\\
&&\left.\left.\left. +P_1.P_4 P_2.P_3 P_2.Z_1 \left(3 \partial_w G^{(3,2,\Delta_\phi,\Delta_\phi)}(u,w)+2 w \partial_w^2 G^{(3,2,\Delta_\phi,\Delta_\phi)}(u,w)\right.\right.\right.\right.\nonumber\\
&&\left.\left.\left.\left. +4 u \left(-\partial_u \partial_w G^{(3,2,\Delta_\phi,\Delta_\phi)}(u,w) \right.\right.\right.\right.\right.\nonumber\\
&&\left.\left.\left.\left.\left. -w \partial_u \partial_w^2 G^{(3,2,\Delta_\phi,\Delta_\phi)}(u,w)+u \partial_u^2 \partial_w G^{(3,2,\Delta_\phi,\Delta_\phi)}(u,w)\right)\right)\right)-2 \left(P_1.P_3\right){}^2 \left(P_2.P_4\right){}^2 w P_2.Z_1\right.\right.\nonumber\\
&&\left.\left. \left(6 \partial_w G^{(3,2,\Delta_\phi,\Delta_\phi)}(u,w) \right.\right.\right.\nonumber\\
&&\left.\left.\left. +w \left(9 \partial_w^2 G^{(3,2,\Delta_\phi,\Delta_\phi)}(u,w)+2 w \partial_w^3 G^{(3,2,\Delta_\phi,\Delta_\phi)}(u,w)\right)\right)+P_1.P_4 P_2.P_3 u \left(P_1.P_4 P_2.P_3 P_2.Z_1\right.\right.\right.\nonumber\\
&&\left.\left.\left.  \left(3 \partial_u G^{(3,2,\Delta_\phi,\Delta_\phi)}(u,w)\right.\right.\right.\right.\nonumber\\
&&\left.\left.\left.\left. +4 u \left(3 \partial_u^2 G^{(3,2,\Delta_\phi,\Delta_\phi)}(u,w)+u \partial_u^3 G^{(3,2,\Delta_\phi,\Delta_\phi)}(u,w)\right)\right)-P_1.P_2 \left(2 P_2.P_4 w P_3.Z_1\right.\right.\right.\right.\nonumber\\
&&\left.\left.\left.\left. \left(\partial_u \partial_w G^{(3,2,\Delta_\phi,\Delta_\phi)}(u,w) \right.\right.\right.\right.\right.\nonumber\\
&&\left.\left.\left.\left.\left. +2 u \partial_u^2 \partial_w G^{(3,2,\Delta_\phi,\Delta_\phi)}(u,w)\right)+P_2.P_3 P_4.Z_1 \left(3 \partial_u G^{(3,2,\Delta_\phi,\Delta_\phi)}(u,w)+4 u \left(3 \partial_u^2 G^{(3,2,\Delta_\phi,\Delta_\phi)}(u,w)\right.\right.\right.\right.\right.\right.\nonumber\\
&&\left.\left.\left.\left.\left.\left. +u \partial_u^3 G^{(3,2,\Delta_\phi,\Delta_\phi)}(u,w)\right)\right)\right)\right)\right) \right)\nonumber\\
&&+\left( \frac{2  \epsilon_{abcde}Z_2^a P_1^b P_3^c P_2^d Z_1^e }{\left(P_1.P_2\right){}^{5/2} P_1.P_3 P_2.P_3 P_2.P_4}\sqrt{\frac{P_2.P_4}{P_1.P_4}}\left(P_3.P_4\right){}^{-\Delta_\phi}\right.\nonumber\\
&&\left. \left(P_1.P_3 P_2.P_4 w \left(\partial_w G^{(3,2,\Delta_\phi,\Delta_\phi)}(u,w)+2 w \partial_w^2 G^{(3,2,\Delta_\phi,\Delta_\phi)}(u,w)\right)-P_1.P_4 P_2.P_3 u \right.\right.\nonumber\\
&&\left.\left. \left(\partial_u G^{(3,2,\Delta_\phi,\Delta_\phi)}(u,w)+2 u \partial_u^2 G^{(3,2,\Delta_\phi,\Delta_\phi)}(u,w)\right)\right) \right)\nonumber\\
&&+\left( \frac{2 \epsilon_{abcde} Z_2^a P_1^b P_4^c P_2^d Z_1^e }{\left(P_1.P_2\right){}^{5/2} \left(P_1.P_4\right){}^2 P_2.P_3 \sqrt{\frac{P_2.P_4}{P_1.P_4}}}\left(P_3.P_4\right){}^{-\Delta_\phi}\right.\nonumber\\
&&\left. \left(P_1.P_4 P_2.P_3 \left(u \left(2 u \partial_u^2 G^{(3,2,\Delta_\phi,\Delta_\phi)}(u,w)-\partial_u G^{(3,2,\Delta_\phi,\Delta_\phi)}(u,w)\right)+G^{(3,2,\Delta_\phi,\Delta_\phi)}(u,w)\right)\right.\right.\nonumber\\
&&\left.\left. -P_1.P_3 P_2.P_4 w \left(3 \partial_w G^{(3,2,\Delta_\phi,\Delta_\phi)}(u,w)+2 w \partial_w^2 G^{(3,2,\Delta_\phi,\Delta_\phi)}(u,w)\right)\right) \right)\nonumber\\
&&\times \frac{1}{(-2)^{\frac{5}{2}+\Delta_\phi}}\nonumber\\
\end{eqnarray}

\section{ Epsilon tensors}\label{epsilontensors}
For polarisations corresponding to $\langle j^x j^+\phi \phi \rangle$, the epsilon tensors are,
\begin{eqnarray}
\epsilon_{abcfe}Z_1^a P_1^b P_3^c P_2^f P_4^e &=& \frac{(a-1) a (z-\bar{z})}{2i} ,\nonumber\\
\epsilon_{abcfe} Z_2^a P_1^b P_3^c P_2^f P_4^e&=& 0, \nonumber\\
\epsilon_{abcfe} Z_2^a P_1^b P_3^c P_2^f Z_1^e&=&\frac{(a-1) a}{i} ,\nonumber\\
\epsilon_{abcfe} Z_2^a P_1^b P_4^c P_2^f Z_1^e&=& \frac{a \bar{z} (a-z)}{i},\nonumber\\
\epsilon_{abcfe} Z_2^a P_1^b P_4^c P_3^f Z_1^e &=&\frac{a \bar{z}-z (a+\bar{z}-1)}{i} ,\nonumber\\
\epsilon_{abcfe} Z_2^a P_2^b P_4^c P_3^f Z_1^e &=& -\frac{(a-1) (\bar{z}-1) (a-z)}{i}.
\end{eqnarray}
For polarisations corresponding to $\langle j^+ j^x\phi \phi \rangle$
\begin{eqnarray}
\epsilon_{abcfe}Z_1^a P_1^b P_3^c P_2^f P_4^e &=& 0 ,\nonumber\\
\epsilon_{abcfe} Z_2^a P_1^b P_3^c P_2^f P_4^e&=& \frac{(a-1) a (z-\bar{z})}{2i}, \nonumber\\
\epsilon_{abcfe} Z_2^a P_1^b P_3^c P_2^f Z_1^e&=& -\frac{(a-1) a}{i},\nonumber\\
\epsilon_{abcfe} Z_2^a P_1^b P_4^c P_2^f Z_1^e&=& \frac{a z (\bar{z}-a)}{i},\nonumber\\
\epsilon_{ancbe} Z_2^a P_1^b P_4^c P_3^f Z_1^e &=&\frac{z (\bar{z}-1)}{i},\nonumber\\
\epsilon_{abcfe} Z_2^a P_2^b P_4^c P_3^f Z_1^e &=& \frac{(a-1) (a (z-1)-z \bar{z}+z)}{i}
\end{eqnarray}
For polarisations corresponding to $\langle j^- j^x \phi \phi\rangle$
\begin{eqnarray}
\epsilon_{abcfe}Z_1^a P_1^b P_3^c P_2^f P_4^e &=& 0 ,\nonumber\\
\epsilon_{abcfe} Z_2^a P_1^b P_3^c P_2^f P_4^e&=&-\frac{(a-1) a (z-\bar{z})}{2i}, \nonumber\\
\epsilon_{abcfe} Z_2^a P_1^b P_3^c P_2^f Z_1^e&=& \frac{(a-1) a}{i},\nonumber\\
\epsilon_{abcfe} Z_2^a P_1^b P_4^c P_2^f Z_1^e&=&-\frac{a \bar{z} (z-a)}{i},\nonumber\\
\epsilon_{abcfe} Z_2^a P_1^b P_4^c P_3^f Z_1^e &=&\frac{(z-1) \bar{z}}{i},\nonumber\\
\epsilon_{abcfe} Z_2^a P_2^b P_4^c P_3^f Z_1^e &=& -\frac{(a-1) (a (\bar{z}-1)-z \bar{z}+\bar{z})}{i}
\end{eqnarray}
 For polarisations corresponding to $\langle j^xj^- \phi \phi \rangle$
\begin{eqnarray}
\epsilon_{abcfe}Z_1^a P_1^b P_3^c P_2^f P_4^e &=& \frac{(a-1) a (z-\bar{z})}{2i}  ,\nonumber\\
\epsilon_{abcfe} Z_2^a P_1^b P_3^c P_2^f P_4^e&=& 0, \nonumber\\
\epsilon_{abcfe} Z_2^a P_1^b P_3^c P_2^f Z_1^e&=& -\frac{(a-1) a}{i},\nonumber\\
\epsilon_{abcfe} Z_2^a P_1^b P_4^c P_2^f Z_1^e&=&-\frac{a z (a-\bar{z})}{i},\nonumber\\
\epsilon_{abcfe} Z_2^a P_1^b P_4^c P_3^f Z_1^e &=&-\frac{(-\bar{z} (a+z)+a z+\bar{z})}{i},\nonumber\\
\epsilon_{abcfe} Z_2^a P_2^b P_4^c P_3^f Z_1^e &=& \frac{(a-1) (z-1) (a-\bar{z})}{i}
\end{eqnarray}                           
\section{Action of $D^t_{[j,\phi]_{\tau_0,l}}$}\label{evenrhsops}
                        In this section we study the effect of differential operators $D^t_{[j,\phi]_{\tau_0,l}}$ on the conformal blocks
\begin{eqnarray}
D^t_{[j,\phi]_{\tau_0,l}}{\cal W}(\Delta_j, \Delta_j, \Delta_\phi, \Delta_\phi) &=& \left( \frac{-1}{l+\Delta_\phi -1} D^t_{11} \sum^{1,0}_L + D^t_{12} \sum^{0,1}_L\right)\nonumber\\
&&\left( \frac{-1}{l+\Delta_\phi -1} D^t_{44} \sum^{1,0}_R + D^t_{43} \sum^{0,1}_R\right) {\cal W}(\Delta_j, \Delta_j, \Delta_\phi, \Delta_\phi)\nonumber\\
\end{eqnarray}  
The kinematics of the four point function is 
\begin{eqnarray}
J(0, Z_1)J(\infty, Z_2) \phi(z, \bar{z})\phi(1,1)\nonumber\\
u=(1-z)(1-\bar{z}),\qquad v=z\bar{z}
\end{eqnarray}

In the subsequent analysis, where we sum over the spins, the effect of the terms subleading in the spin $l$ can be ignored. 
\begin{eqnarray}
&& D^t_{12} \sum^{0,1}_L D^t_{43} \sum^{0,1}_R {\cal W}(\Delta_j, \Delta_j, \Delta_\phi, \Delta_\phi)\nonumber\\
&&=\frac{1}{4 \left(P_1.P_2\right){}^4}\left(\frac{P_1.P_2}{P_1.P_3}\right){}^{\frac{1}{2} (-\Delta_j+\Delta_\phi+3)} \left(P_1.P_4\right){}^{\frac{1}{2} (-\Delta_j-\Delta_\phi-1)} \left(P_2.P_3\right){}^{\frac{1}{2} (-\Delta_j-\Delta_\phi-1)}\nonumber\\
&& \left(\frac{P_2.P_4}{P_1.P_2}\right){}^{\frac{1}{2} (\Delta_j-\Delta_\phi-3)} \left(P_1.P_2 P_1.P_4 P_3.Z_1 \left(P_1.P_2 \left(P_2.P_4 P_3.Z_2 \left((-\Delta_j+\Delta_\phi+1)^2 G(v,u) \right.\right.\right.\right.\nonumber\\ 
&&\left.\left.\left.\left. -2 \left(u (\Delta_j-\Delta_\phi-1) \partial_u G(v,u)+2 v \left(\partial_v G(v,u)+v \partial_v^2 G(v,u)+u \partial_u \partial_v G(v,u)\right)\right)\right)+P_2.P_3 P_4.Z_2 \right.\right.\right.\nonumber\\
&&\left.\left.\left. \left(G(v,u) (-\Delta_j+\Delta_\phi+1)^2+4 \left(\partial_v^2 G(v,u) v^2+(-\Delta_j+\Delta_\phi+2) \partial_v G(v,u) v+u \left((-\Delta_j+\Delta_\phi+2) \right.\right.\right.\right.\right.\right.\nonumber\\
&& \left.\left.\left.\left.\left.\left. \partial_u G(v,u)+2 v \partial_u \partial_v G(v,u)+u \partial_u^2 G(v,u)\right)\right)\right)\right)-2 P_1.Z_2 P_2.P_3 P_2.P_4 \left(G(v,u) (-\Delta_j+\Delta_\phi+1)^2 \right.\right.\right.\nonumber\\
&&\left.\left.\left. +2 v \partial_v G(v,u) (-\Delta_j+\Delta_\phi+1)+u \left((-3 \Delta_j+3 \Delta_\phi+5) \partial_u G(v,u)+2 \left(v \partial_u \partial_v G(v,u)+u \partial_u^2G(v,u)\right)\right)\right)\right) \right.\nonumber\\
&&\left. +P_1.P_3 \left(P_1.P_2 P_4.Z_1 \left(2 P_1.Z_2 P_2.P_3 P_2.P_4 \left(2 u v \partial_u \partial_v G(v,u)-(\Delta_j-\Delta_\phi-1) \left((\Delta_j-\Delta_\phi-1) G(v,u) \right.\right.\right.\right.\right.\nonumber\\
&&\left.\left.\left.\left.\left. +2 v \partial_v G(v,u)-u \partial_u G(v,u)\right)\right)+P_1.P_2 \left(P_2.P_4 P_3.Z_2 \left(G(v,u) (-\Delta_j+\Delta_\phi+1)^2 \right.\right.\right.\right.\right.\nonumber\\
&&\left.\left.\left.\left.\left. +4 v \left((\Delta_j-\Delta_\phi) \partial_v G(v,u)+v \partial_v^2 G(v,u)\right)\right)+P_2.P_3 P_4.Z_2 \left((-\Delta_j+\Delta_\phi+1)^2 G(v,u) \right.\right.\right.\right.\right.\nonumber\\
&&\left.\left.\left.\left.\left. -2 \left(u (\Delta_j-\Delta_\phi-1) \partial_u G(v,u)+2 v \left(\partial_v G(v,u)+v \partial_v^2 G(v,u)+u \partial_u \partial_v G(v,u)\right)\right)\right)\right)\right) \right.\right.\nonumber\\
&&\left.\left. -2 P_1.P_4 \left(2 P_1.Z_2 P_2.P_3 P_2.P_4 P_2.Z_1 \left(-\partial_u^2G(v,u) u^2-(\Delta_j-\Delta_\phi-1) \left((\Delta_j-\Delta_\phi) G(v,u) \right.\right.\right.\right.\right.\nonumber\\
&&\left.\left.\left.\left.\left. -2 u \partial_u G(v,u)\right)\right)+P_1.P_2 \left(P_2.P_4 \left(P_2.Z_1 P_3.Z_2 \left((\Delta_j-\Delta_\phi-1) \left((\Delta_j-\Delta_\phi-1) G(v,u) \right.\right.\right.\right.\right.\right.\right.\nonumber\\
&&\left.\left.\left.\left.\left.\left.\left. +2 v \partial_v G(v,u)-u \partial_u G(v,u)\right)-2 u v \partial_u \partial_v G(v,u)\right)-2 P_2.P_3 Z_1.Z_2 \left((-\Delta_j+\Delta_\phi+1) G(v,u) \right.\right.\right.\right.\right.\right.\nonumber\\
&&\left.\left.\left.\left.\left.\left. +u \partial_u G(v,u)\right)\right)+P_2.P_3 P_2.Z_1 P_4.Z_2 \left(G(v,u) (-\Delta_j+\Delta_\phi+1)^2+2 v \partial_v G(v,u) (-\Delta_j+\Delta_\phi+1) \right.\right.\right.\right.\right.\nonumber\\
&&\left.\left.\left.\left.\left. +u \left((-3 \Delta_j+3 \Delta_\phi+5) \partial_u G(v,u)+2 \left(v \partial_u \partial_v G(v,u)+u \partial_u^2G(v,u)\right)\right)\right)\right)\right)\right)\right)\nonumber\\
\end{eqnarray} 
where $G(v,u)\sim G^{(\Delta_j,\Delta_j,\Delta_{\phi}+1,\Delta_{\phi}+1)} (v,u)$.                      
For conserved currents $\Delta_j=2$ in  $d=3$. 
\section{Action of $D^{11}_m$}\label{mixedrhsops}
\begin{eqnarray}
&&\tilde{D}^t_1 D^t_{43} \sum^{0,1}_R{\cal W}_{\tilde{\mathcal{O}}}^t(\Delta_j, \Delta_j, \Delta_\phi, \Delta_\phi)\nonumber\\
&&=\frac{1}{8 \left(P_1.P_3\right){}^2 \left(P_2.P_4\right){}^3 P_3.P_4}\left(\frac{P_1.P_2}{P_1.P_3}\right){}^{\frac{1}{2} (-\Delta_j+\Delta_\phi-1)} \left(P_1.P_4\right){}^{\frac{1}{2} (-\Delta_j-\Delta_\phi)} \left(P_2.P_3\right){}^{\frac{1}{2} (-\Delta_j-\Delta_\phi-1)} \nonumber\\
&&\left(\frac{P_2.P_4}{P_1.P_2}\right){}^{\frac{1}{2} (\Delta_j-\Delta_\phi+2)} \epsilon_{abcfe}P^a_1P^b_2P^c_3P^f_4Z^e_1 \nonumber\\
&&\left(P_3.P_4 \left(P_2.P_4 P_3.Z_2 \left((\Delta_j-\Delta_\phi) G(v,u) (-\Delta_j+\Delta_\phi+1)^2+2 \left(2 (-\Delta_j+\Delta_\phi+6) \partial_v^2 G(v,u) v^2 \right.\right.\right.\right.\nonumber\\
&&\left.\left.\left.\left. -v \left((\Delta_j-\Delta_\phi)^2-3\right) \partial_v G(v,u)-u (2 \Delta_j-2 \Delta_\phi-3) (\Delta_j-\Delta_\phi-1) \partial_u G(v,u) \right.\right.\right.\right.\nonumber\\
&&\left.\left.\left.\left. +2 \left(2 \partial_v^3 G(v,u) v^3+u \left(u (\Delta_j-\Delta_\phi-1) \partial_u^2 G(v,u)+v \left(5 \partial_u \partial_v G(v,u) \right.\right.\right.\right.\right.\right.\right.\nonumber\\
&&\left.\left.\left.\left.\left.\left.\left. +4 v \partial_u \partial_v^2 G(v,u)+2 u \partial_u^2 \partial_v G(v,u)\right)\right)\right)\right)\right)+P_2.P_3 P_4.Z_2 \left((\Delta_j-\Delta_\phi-2) (\Delta_j-\Delta_\phi-1)\right.\right.\right.\nonumber\\
&&\left.\left.\left. (\Delta_j-\Delta_\phi) G(v,u)-2 \left(6 (-\Delta_j+\Delta_\phi+3) \partial_v^2 G(v,u) v^2+3 (-\Delta_j+\Delta_\phi+2)^2 \partial_v G(v,u) v \right.\right.\right.\right.\nonumber\\
&&\left.\left.\left.\left. +3 u (-\Delta_j+\Delta_\phi+2)^2 \partial_u G(v,u)+2 \left(2 \partial_v^3 G(v,u) v^3+u \left(2 \partial_u^3 G(v,u) u^2+3 (-\Delta_j+\Delta_\phi+3)\right.\right.\right.\right.\right.\right.\nonumber\\
&& \left.\left.\left.\left.\left.\left. \partial_u^2 G(v,u) u+6 v (-\Delta_j+\Delta_\phi+3) \partial_u \partial_v G(v,u)+6 v \left(v \partial_u \partial_v^2 G(v,u)+u \partial_u^2 \partial_v G(v,u)\right)\right)\right)\right)\right)\right) \right.\nonumber\\
&&\left. \left(P_1.P_2\right){}^2+P_2.P_4 \left(2 u P_1.P_3 \left(P_2.P_4 P_3.Z_2 \left((\Delta_j-\Delta_\phi-1) (2 \Delta_j-2 \Delta_\phi-3) \partial_u G(v,u)\right.\right.\right.\right.\nonumber\\
&&\left.\left.\left.\left. +2 \left(v (2 \Delta_j-2 \Delta_\phi-3) \partial_u \partial_v G(v,u)+u (-\Delta_j+\Delta_\phi+1) \partial_u^2 G(v,u)-2 u v \partial_u^2 \partial_v G(v,u)\right)\right)\right.\right.\right.\nonumber\\
&&\left.\left.\left. +P_2.P_3 P_4.Z_2 \left((\Delta_j-\Delta_\phi-2) (2 \Delta_j-2 \Delta_\phi-3) \partial_u G(v,u)+2 v (-2 \Delta_j+2 \Delta_\phi+3) \partial_u \partial_v G(v,u)\right.\right.\right.\right.\nonumber\\
&&\left.\left.\left.\left. +2 u \left((-3 \Delta_j+3 \Delta_\phi+7) \partial_u^2 G(v,u)+2 \left(v \partial_u^2 \partial_v G(v,u)+u \partial_u^3 G(v,u)\right)\right)\right)\right)\right.\right.\nonumber\\
&&\left.\left. +P_1.Z_2 P_2.P_3 P_3.P_4 \left(2 \left(2 (-2 \Delta_j+2 \Delta_\phi+1) \partial_v^2 G(v,u) v^2+(2 \Delta_j-2 \Delta_\phi-3) (2 \Delta_j-2 \Delta_\phi-1)\right.\right.\right.\right. \nonumber\\
&&\left.\left.\left.\left. \partial_v G(v,u) v+u \left(\left(5 \Delta_j^2-(10 \Delta_\phi+13) \Delta_j+\Delta_\phi (5 \Delta_\phi+13)+9\right) \partial_u G(v,u)\right.\right.\right.\right.\right.\nonumber\\
&&\left.\left.\left.\left.\left.+2 \left(3 v (-2 \Delta_j+2 \Delta_\phi+3) \partial_u \partial_v G(v,u)+2 \left(\partial_u \partial_v^2 G(v,u) v^2 \right.\right.\right.\right.\right.\right.\right.\nonumber\\
&&\left.\left.\left.\left.\left.\left.\left. +u \left(2 (-\Delta_j+\Delta_\phi+2) \partial_u^2 G(v,u)+2 v \partial_u^2 \partial_v G(v,u)+u \partial_u^3 G(v,u)\right)\right)\right)\right)\right)\right.\right.\right.\nonumber\\
&&\left.\left.\left. -(2 \Delta_j-2 \Delta_\phi-1) (\Delta_j-\Delta_\phi-1) (\Delta_j-\Delta_\phi) G(v,u)\right)\right) P_1.P_2+2 u P_1.P_3 P_1.Z_2 P_2.P_3 \left(P_2.P_4\right){}^2\right.\nonumber\\
&&\left. \left((2 \Delta_j-2 \Delta_\phi-3) \left((-2 \Delta_j+2 \Delta_\phi+1) \partial_u G(v,u)+4 u \partial_u^2 G(v,u)\right)-4 u^2 \partial_u^3 G(v,u)\right)\right)\frac{1}{(-2)^{\frac{5}{2}+\Delta_\phi}}\nonumber\\
&&-\frac{1}{4 \left(P_1.P_3\right){}^2 P_2.P_4 P_3.P_4}\left(\frac{P_1.P_2}{P_1.P_3}\right){}^{\frac{1}{2} (-\Delta_j+\Delta_\phi-1)} \left(P_1.P_4\right){}^{\frac{1}{2} (-\Delta_j-\Delta_\phi)} \left(P_2.P_3\right){}^{\frac{1}{2} (-\Delta_j-\Delta_\phi+1)} \nonumber\\
&&\left(\frac{P_2.P_4}{P_1.P_2}\right){}^{\frac{\Delta_j-\Delta_\phi}{2}} \epsilon_{abcfe} P_1^aP_3^bP_4^cZ_1^fZ_2^e \left(P_1.P_2 P_3.P_4 \left(4 \left(u^2 \partial_u^2 G(v,u)+v^2  \partial_v^2 G(v,u)\right.\right.\right.\nonumber\\
&&\left.\left.\left. +2 u v \partial_u \partial_v G(v,u)\right)+2 v (-2 \Delta_j+2 \Delta_\phi+3)  \partial_v G(v,u)+2 u (-2 \Delta_j+2 \Delta_\phi+3) \partial_u  G(v,u)\right.\right.\nonumber\\
&&\left.\left. +(\Delta_j-\Delta_\phi-1) (\Delta_j-\Delta_\phi) G(u,v)\right)-2 P_1.P_3 P_2.P_4 u \left((-2 \Delta_j+2 \Delta_\phi+3) \partial_u  G(v,u)\right.\right.\nonumber\\
&&\left.\left.+2 u \partial_u^2 G(v,u)\right)\right)\frac{1}{(-2)^{\frac{5}{2}+\Delta_\phi}}
\end{eqnarray}
where $G(v,u) \sim G^{(\Delta_j,\Delta_j,\Delta_\phi+1,\Delta_\phi)}(v,u)$

\providecommand{\href}[2]{#2}\begingroup\raggedright\endgroup

 \end{document}